\documentclass[%
 aip,
 amsmath,amssymb,
 reprint,%
]{revtex4-1}

\usepackage{graphicx}% Include figure files
\usepackage{dcolumn}% Align table columns on decimal point
\usepackage{bm}% bold math
\usepackage{ragged2e}
\usepackage{booktabs}
\usepackage{geometry}
\usepackage{graphicx}
\usepackage{multirow}
\usepackage{tabularx}
\usepackage{graphicx}
\usepackage{caption}
\usepackage{dcolumn}
\usepackage{bm}
\usepackage{hyperref}
\usepackage{natbib}
\usepackage{amssymb}
\usepackage{amsmath}
\usepackage{stackengine}
\usepackage{pgfplots}
\usepackage{adjustbox}
\usepackage{makecell, multirow, tabularx}
\usepackage{array}
\usepackage{booktabs}
\usepackage{mathtools}

\usepackage{float}
\usepackage{caption}
\usepackage{subcaption}
\pgfplotsset{compat=1.18}

\usepackage[utf8]{inputenc}
\usepackage[T1]{fontenc}
\usepackage{mathptmx}
\usepackage{etoolbox}

\makeatletter
\def\@email#1#2{%
 \endgroup
 \patchcmd{\titleblock@produce}
  {\frontmatter@RRAPformat}
  {\frontmatter@RRAPformat{\produce@RRAP{*#1\href{mailto:#2}{#2}}}\frontmatter@RRAPformat}
  {}{}
}%
\makeatother
\begin{document}

\preprint{AIP/123-QED}

\title{Causality Analysis of COVID-19 Induced Crashes in Stock and Commodity Markets: A Topological Perspective}
%\title{COVID-19 induced crashes in stock and commodity markets: Analysis using TDA and Granger-Causality\\}
% Force line breaks with \\

\author{Buddha Nath Sharma}%
\email{bnsharma09@yahoo.com}
\affiliation{Department of Physics, National Institute of Technology Sikkim, Sikkim, India-737139.%\\This line break forced with \textbackslash\textbackslash
}%
\author{Anish Rai}
\email{anishrai412@gmail.com}
\affiliation{AlgoLabs, Chennai Mathematical Institute, India-603103.}
\affiliation{Department of Physics, National Institute of Technology Sikkim, Sikkim, India-737139.}%Lines break automatically or can be forced with \\

\author{Salam Rabindrajit Luwang}
\email{salamrabindrajit@gmail.com}
\affiliation{Department of Physics, National Institute of Technology Sikkim, Sikkim, India-737139.}

\author{Md. Nurujjaman}
\email{md.nurujjaman@nitsikkim.ac.in}
\affiliation{Department of Physics, National Institute of Technology Sikkim, Sikkim, India-737139.}
	
\author{Sushovan Majhi}
\email{s.majhi@gwu.edu}
\affiliation{Data Science Program, George Washington University, Washington, DC, USA, 20052.}
\homepage{https://www.smajhi.com}
%\affiliation{%
%Second institution and/or address%\\This line break forced% with \\
%}%


\begin{abstract}

The paper presents a comprehensive causality analysis of the US stock and commodity markets during the COVID-19 crash. The dynamics of different sectors are also compared. We use Topological Data Analysis (TDA) on multidimensional time-series to identify crashes in stock and commodity markets. The Wasserstein Distance (\emph{WD}) shows distinct spikes signaling the crash for both stock and commodity markets. We then compare the persistence diagrams of stock and commodity markets using the \emph{$WD$} metric. A significant spike in the $WD$ between stock and commodity markets is observed during the crisis, suggesting significant topological differences between the markets. Similar spikes are observed between the sectors of the US market as well. Spikes obtained may be due to either a difference in the magnitude of crashes in the two markets (or sectors), or from the temporal lag between the two markets suggesting information flow. We study the Granger-causality between stock and commodity markets and also between different sectors. The results show a bidirectional Granger-causality between commodity and stock during the crash period, demonstrating the greater interdependence of financial markets during the crash. However, the overall analysis shows that the causal direction is from stock to commodity. A pairwise Granger-causal analysis between US sectors is also conducted. There is a significant increase in the interdependence between the sectors during the crash period. TDA combined with Granger-causality effectively analyzes the interdependence and sensitivity of different markets and sectors.
  
%The results establish superiority of the stock market as compared to the commodity market during the period. \textbf{The H0 first difference time-series of $WD_1$ and $WD_2$ also show Granger-causality from Stock to commodity during the period of crash(2020-01-01 to 2020-06-01)}
  
\end{abstract}

\maketitle

\section{Introduction}
\label{Introduction}

The financial market reflects the economic health and stability of a country, serving as the platform for investment and growth~\cite{marques2013does,wong2011development}. However, they are susceptible to rare and catastrophic events -- such as the dot-com bubble of the early 2000s, the 2008 global financial crisis, and the COVID-19 pandemic, leading to extreme crashes. These events changed the dynamics of the market leading to extreme losses of investors and traders~\cite{sornette2003critical,voit2003statistical}. In particular, the recent crash due to the COVID-19 pandemic severely impacted the world economy, unlike previous devastating crashes~\cite{qureshi2022covid}, resulting in extreme losses and destabilization of financial systems. Its impacts were transmitted between various financial instruments, including stocks and commodities, and within various sectors resulting in increased interconnectedness and interdependence during the crash period~\cite{adekoya2021covid,chirilua2022connectedness,costa2022sectoral}. Therefore, it is crucial to detect the crash and analyze the interdependence between various financial instruments in and around the crash period.

Financial markets exhibit complex dynamics\cite{zhang2017nonlinear,tenreiro2013complex}. Hence, the detection of market crashes is challenging. Traditional methods, such as the Fourier transform and its variants~\cite{bracewell1966fourier,ransom2002fourier,xu2010antileakage}, as well as least squares spectral analysis~\cite{ghaderpour2018least}, do not detect abrupt shifts in nonlinear and nonstationary time series and, consequently, lack robustness. Few improved methods such as wavelet transform~\cite{foster1996wavelets} and Hilbert-Huang transform~\cite{mahata2020identification,rai2023detection,rai2022sentiment,rai2022statistical} that successfully handle nonlinear data can analyze only one time-series at a time. Topological data analysis (TDA) has proven to overcome such limitations by capturing the collective dynamics and inter-relationships of the broader markets when analyzing multiple time series simultaneously~\cite{rai2024identifying}. TDA can also quantify topological structural changes proving it to be the most suitable method for the detection of crashes~\cite{akingbade2024topological}. Further, the method is robust to small perturbations, making TDA ideal for analyzing financial markets, which are frequently influenced by noise traders. Hence, considering these advantages, in this study, we utilize TDA to analyze financial markets and detect crashes during COVID-19 pandemic period.

%Analyzing the dynamics of the financial market is challenging due to its complex, nonlinear behavior. Traditional methods such as Fourier transform and its variants~\cite{burns1969pulsar,bracewell1966fourier,ransom2002fourier,gabor1946theory,xu2010antileakage,guo2015high} usually fail to analyze the abrupt shifts in non-linear and non-stationary time series and, consequently, lack robustness. Least-squares spectral analysis methods are also used to analyze trends and/or datum shifts~\cite{vanivcek1969approximate} but assume linearity. Further, few improved methods such as wavelet transform techniques~\cite{morlet1983sampling,grossmann1985transforms,foster1996wavelets,torrence1998practical,ghaderpour2017least}, Hilbert-Huang Transform (HHT)~\cite{mahata2020identification,rai2023detection,rai2022sentiment,rai2022statistical} and the heterogeneous economic time model~\cite{soloviev2011heisenberg,kuzu2022model}---that successfully handle the complexity can analyze only $1$-dimensional time-series at once. These reasons are of concern when understanding the collective financial and sectoral markets. Topological data analysis (TDA) has proven to overcome such limitations, which makes it appropriate for studying complex systems such as financial markets.\\
TDA has been applied in a number of fields for analyzing noisy multi-dimensional datasets~\cite{kramar2013persistence,seversky2016time,nicolau2011topology,lee2017quantifying,syed2021using}. TDA has also been used extensively in financial markets. These include detection of crashes in stock and cryptocurrency market~\cite{aguilar2020topology,yen2021understanding,yen2021using,guo2021analysis,guo2022risk}, early warning signals for stock market crashes~\cite{gidea2017topology,gidea2018topological,gidea2020topological}, portfolio optimization~\cite{goel2020topological} etc. Some of the important developments regarding the applications of TDA in finance are discussed in Ref.~\citenum{rai2024identifying}. In this study, we detect crash due to COVID-19 in stock and commodity markets using TDA. We also perform a direct topological comparison of the two markets. A similar analysis is conducted for the major sectors of US stock market.

Granger-causality is an established technique to study the interactions between two time-series~\cite{shojaie2022Granger}. It has been used to analyze the interdependence between markets (or sectors)~\cite{kang2013linkage,hong2009granger}. However, the studies so far have focused on the Granger-causal relations between the return and variance of the individual financial market time-series ~\cite{kang2013linkage,patel2013causal,fernandez2014linear,choudhry2015relationship}. TDA, however, provides the technique to incorporate multiple time-series of each market (or sector). In addition to that, TDA effectively captures the significant periods such as bubbles or crashes, and distinguishes them from the normal periods as illustrated in Ref.~\citenum{akingbade2024topological}. Therefore, the application of Granger-causality in addition to TDA provides a comprehensive analysis of the interdependence between different markets (or sectors).

% There have been many studies regarding the Granger-causal relations between different stocks(or sectors)/commodities etc,. However, no such studies have been conducted by considering multiple constituents from each market/sector at once. TDA overcomes this limitation by incorporating multiple time-series to produce a time-series of topological metric such as Wasserstein Distance $\mathcal{WD}$. The study of Granger-causal relations between the $\mathcal{WD}$ time-series of the different markets(or sectors) gives a complete picture of the interdependence, spillovers and information flow between different markets(or sectors).\\
%\textcolor{red}{Add our motivation behind using TDA in this paper.}
%Although the commodity and stock marThe financialization of commodity markets has increased the correlation between commodity and stock markets [cite]. Hence, the interdependence between the two markets is an important area of study for investors seeking portfolio diversification. Further, the COVID-19 pandemic had a direct impact on the 
Our work aspires to analyze crashes in markets and sectors due to COVID-19 and their interdependence in and around that period. We consider the stock and commodity markets as the financialization of commodity market has increased the correlation between commodity and stock markets~\cite{kang2023financialization}. Moreover, the COVID-19 pandemic and the measures taken to contain it directly impacted the supply chains and production processes of major commodities~\cite{bank2020shock}. Hence, the interdependence between the two markets is an important area of study for investors. Moreover, we have chosen commodity and stock markets since these markets have the same trading days during this period---a feature that mitigates information loss. Moreover, the different sectors of the US market were impacted differently by the  crash, although all sectors experienced a crash~\cite{mazur2021covid}. Therefore, the study of interdependence between different sectors is of utmost importance.

In this work, we detect the crashes in stock and commodity markets due to COVID-19, using techniques from TDA. Further, we study the topological comparison between the stock and commodity markets during the same period. Lastly, we divide the time period into \emph{pre-crash}, \emph{crash}, and \emph{post-crash} periods to study the Granger-causality between the two markets during these periods.  Subsequently, we conduct a similar analysis for major sectors of the US stock market.

The structure of our paper is as follows.  Sec.~\ref{sec:methodology} explains the methodology employed in the paper and Sec.~\ref{Data} contains the data analyzed. The approach used to implement Granger-causality with TDA is explained in Sec.~\ref{gc_appraoch}. In Sec.~\ref{result} we have discussed our results and Sec.~\ref{con} contains the conclusion of the work.

\section{Methodology}
\label{sec:methodology}
The work uses the tools from topological data analysis (TDA) to study the multidimensional time-series of a market (or sector) at once. TDA is used to identify market crashes in different markets (or sectors). Then, for the time-series generated from TDA, we use the statistical techniques of Granger-causality to study the influence between the markets (or sectors) during different time periods. The following sections present a brief introduction to the two methods used in the study.\\
\subsection{Topological Data Analysis}
\label{sec:TDA}

Topological Data Analysis (TDA) facilitates the extraction of robust topological features from noisy data sets. It is motivated to infer robust qualitative and quantitative information about the structure of the data through its topology and geometry~\cite{chazal2021introduction}. One significant advantage of TDA is its robustness; the output remains unchanged in the presence of small perturbations~\cite{gidea2018topological}. Persistent Homology (PH), a potent technique in TDA, is used in this study. It offers a quantitative explanation of how the connectivity and structure of data evolve as the resolution ($\varepsilon$) changes~\cite{gold2023algorithm}.

Sec.~\ref{Rips}--~\ref{WassersteinD} present the definitions and notations of persistent homology used in this paper.

\subsubsection{Vietoris--Rips Complexes}
\label{Rips}

To utilize persistent homology (PH), we initiate with a point cloud dataset, $X$, which is depicted as a collection of points within the $d$-dimensional Euclidean space $\mathbb{R}^d$. For a given scale $\varepsilon > 0$, we relate a simplicial complex to the data set through the Vietoris--Rips complex, also known as the Rips complex, denoted as $\mathcal{R}(X, \varepsilon))$. It effectively captures the connectivity among data points at varying $\varepsilon$ values. Although the Rips complex can be generalized, we elucidate the concept here specifically for Euclidean point clouds. A subset $\{x_0, x_1, \ldots, x_k\}$ of $X$ consisting of $(k+1)$ points forms a $k$-simplex in the Rips complex if every pairwise distance $\|x_i - x_j\|$ is no greater than $\varepsilon$~\cite{gidea2018topological, souto2023topological, akingbade2024topological}.

The Rips complexes constitute a filtration. This means that as we increase the value of $\varepsilon$, the simplicial complexes include a greater number of simplices~\cite{akingbade2024topological,chazal2021introduction}.
For each complex in the filtration, we can compute its $k$-dimensional homology group $H_k(\mathcal{R}(X, \varepsilon))$ using coefficients in a chosen field. These homology groups identify topological features of different dimensions present in the dataset at each scale $\varepsilon$. For instance, the $0$-dimensional homology group $H_0(\mathcal{R}(X, \varepsilon))$ represents the connected components, the $1$-dimensional homology group $H_1(\mathcal{R}(X, \varepsilon))$ captures independent loops and so on.

% As the $\varepsilon$ increases, some topological features of different dimensions (connected components for 0-dimension, loops for 1-dimension and so on) are born and some topological features die. The topology of the point cloud is understood by studying the persistence of these features with increase in $\varepsilon$.

For two resolutions $0<\varepsilon_1\leq\varepsilon_2$, the emergence and disappearance of $k$-dimensional topological features can be analyzed through the homomorphism induced by the natural inclusion~\cite{Dey_Wang_2022}:  
\[
H_k(\mathcal{R}(X, \varepsilon_1))\xhookrightarrow{\quad\iota_*\quad} H_k(\mathcal{R}(X, \varepsilon_2)),
\]
which maps the $k$-th homology groups. A non-trivial homology class in $H_k(\mathcal{R}(X, \varepsilon_1))$ may vanish in $H_k(\mathcal{R}(X, \varepsilon_2))$, while new homology classes may emerge in the latter that are not inherited through the inclusion from the former.

\begin{figure}[h!]
    \centering
    \includegraphics[width=0.9\linewidth]{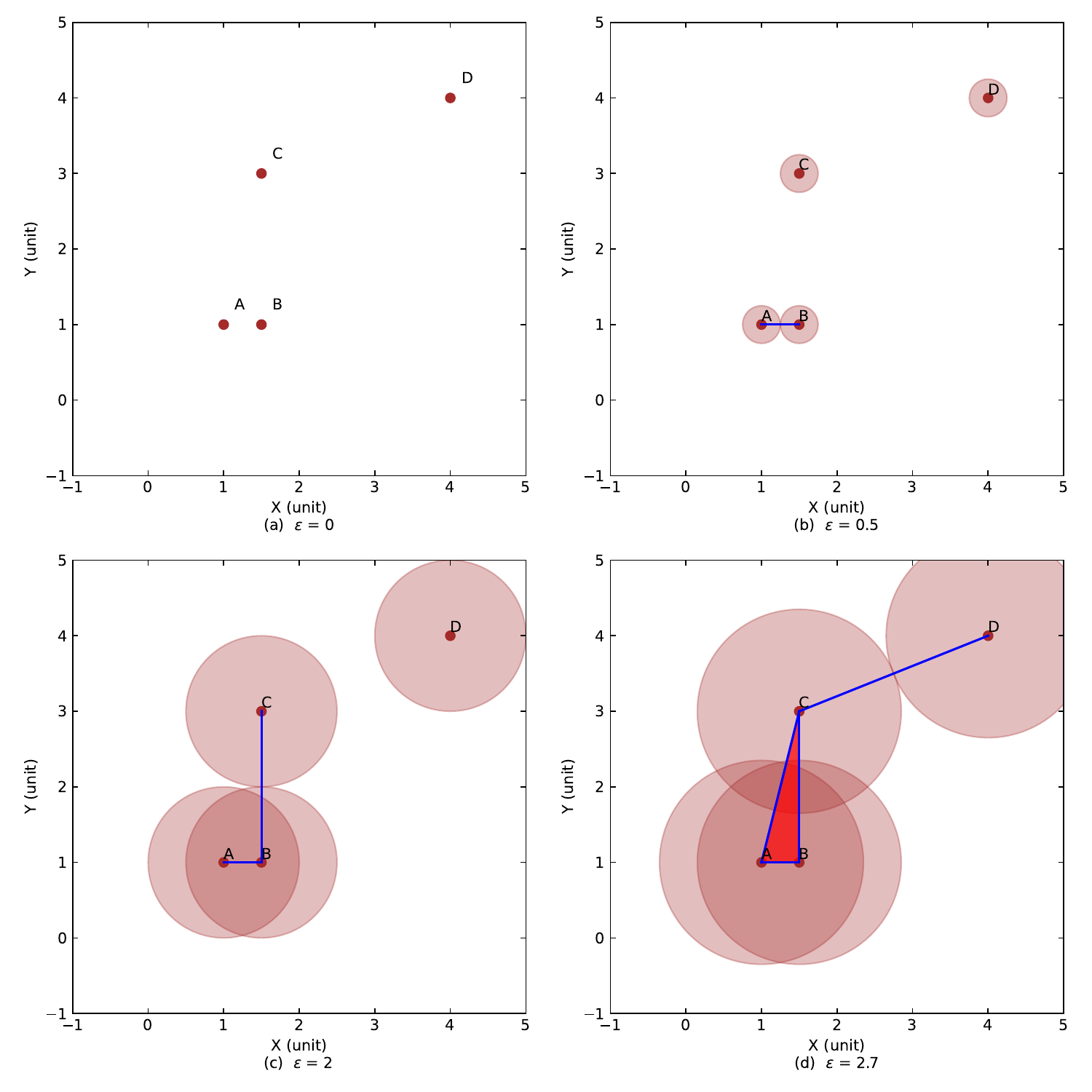}
    \caption{\justifying \small
    \textit{Figures depict the Rips complex for the example dataset at various resolutions($\varepsilon$). Figs. (a),(b),(c) and (d) represent the Rips complex at $\varepsilon= 0,0.5,2$ and $2.7$ respectively.}}
    \label{fig:Rips}
\end{figure}

We illustrate the concept by considering the following set of four data points:  
$S = \{A(1, 1), B(1.5, 1), C(1.5, 3), D(4, 4)\}
\subset\mathbb{R}^2$.
The Rips complexes for this point set are constructed at various $\varepsilon$ and shown in Fig.~\ref{fig:Rips}. The points remain isolated till $\varepsilon = 0.49$ as shown in Fig.~\ref{fig:Rips}(a). At $\varepsilon = 0.5$ points $A$ and $B$ become connected and hence they are connected by line $AB$ as shown in Fig.~\ref{fig:Rips}(b). At $\varepsilon = 2$, points $B$ and $C$ also become connected; therefore, they are connected by line $BC$ as shown in Fig.~\ref{fig:Rips}(c). At $\varepsilon = 2.7$, $C$ and $D$ also become connected, so a line $CD$ is formed between the points as shown in Fig.~\ref{fig:Rips}(d). At this point, all points are connected to form a single connected component. We also observe a loop $ABC$ as the three points form a pairwise connection with each other.

%Such a study is facilitated by \emph{Persistence Diagrams}, which is described in the following section.

%For each complex in the filtration, we can compute its $k$-dimensional homology $H_k(\mathcal{R}(X, \varepsilon))$ with coefficients in some field. The homology groups capture the different dimensional topological features present in the dataset at each $\varepsilon$. For example, the $0$-dimensional homology group $H_0(\mathcal{R}(X, \varepsilon))$ corresponds to the connected components, the $1$-dimensional homology group $H_1(\mathcal{R}(X, \varepsilon))$ corresponds to the independent loops, the $2$-dimensional homology group $H_2(\mathcal{R}(X, \varepsilon))$ corresponds to the independent voids, and so on.

%For two resolutions $0<\varepsilon_1\leq\varepsilon_2$, we can identify the birth and death of $k$-dimensional topological features by examining the homomorphism induced by natural inclusion~\cite{Dey_Wang_2022} $$H_k(\mathcal{R}(X, \varepsilon_1))\xhookrightarrow{\quad\iota_*\quad} H_k(\mathcal{R}(X, \varepsilon_2))$$ between the $k$-th homology groups. A non-trivial homology class in $H_k(\mathcal{R}(X, \varepsilon_1))$ may disappear in $H_k(\mathcal{R}(X, \varepsilon_2))$; likewise, a new homology class may appear in the latter that is not \emph{carried} by the inclusion from the former. 

\subsubsection{Persistence Diagrams}
\label{persistenceD}
The importance of topological features is determined by their persistence. Features that persist for an extended resolution range are considered `significant,' while those that disappear quickly are termed `noisy'~\cite{guo2020empirical}. Persistence diagram summarizes the birth and death of these features in a two-dimensional plot. This diagram plots birth and death coordinates, represented by two-dimensional points with their multiplicities. According to the stability theorem of persistence diagrams, it remains stable under small and potentially irregular perturbations~\cite{cohen2005stability}.
\begin{figure}[h!]
    \centering
    \includegraphics[width=0.9\linewidth]{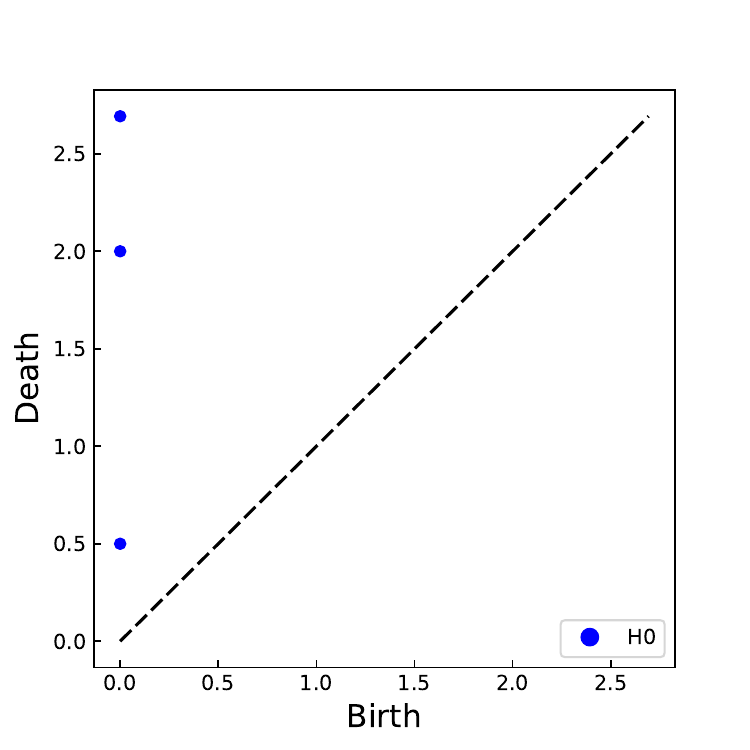}
    \caption{\justifying \small
    \textit{Presistence diagram for $0$-dimensional homology group for the example dataset}}
    \label{fig:PD_dummy}
\end{figure}

%The significance of the topological features is understood through their persistence. The features that persist for a long resolution range are called `significant' and the ones dying within a short resolution range are termed `noisy' features~\cite{guo2020empirical}. The summary of the birth and death of these features with change in resolution is preserved in a two-dimensional figure named \emph{persistence diagram}. It is a plot between the birth and death coordinates, consisting of two-dimensional points with their multiplicities. As per the stability theorem of persistence diagrams, it is stable under possibly irregular perturbations of small amplitude~\cite{cohen2005stability}.

The \emph{persistence diagram} $\mathcal{PD}_k$ for the $k$-dimensional homology comprises:
\begin{itemize}
    \item For each $k$-dimensional feature that takes birth at scale $b$ and dies at scale $d$, a point $p = p(b, d) \in \mathbb{R}^2$ along with its multiplicity $\mu$;
    \item $\mathcal{PD}_k$ also includes all points on the positive diagonal of $\mathbb{R}^2$; these points denote trivial homology generators that emerge and vanish instantaneously at every level, each point on the diagonal having infinite multiplicity.
\end{itemize}
Although the persistence diagrams (\emph{PDs}) effectively summarize the birth and death of topological features, derivation of quantitative inferences from the (\emph{PDs}) requires the endowment of a metric, such as \emph{Wasserstein Distance}, which is described in the following section.\\
The \emph{PD} corresponding to the $0$-dimensional homology group for Fig.~\ref{fig:Rips} is shown in Fig.~\ref{fig:PD_dummy}. Although, all the connected components are born at $\varepsilon=0$, they merge and hence die at $0.5, 2.0$, and $2.7$ respectively as shown by the blue dots. From $\varepsilon$=$2.7$, only one connected component remains. 

\subsubsection{Wasserstein Distance}
\label{WassersteinD}
The space of persistence diagrams (\emph{PDs}) can be equipped with a metric structure. A commonly employed metric is the degree $p$-Wasserstein distance~\cite{gidea2018topological}, denoted $WD_p(\mathcal{PD}_k^1, \mathcal{PD}_k^2)$, where $\mathcal{PD}_k^1$ and $\mathcal{PD}_k^2$ are two \emph{PDs} in dimension $k$. The distance $WD_p(\mathcal{PD}_k^1, \mathcal{PD}_k^2)$ is defined as:
\begin{equation*}
\resizebox{1\columnwidth}{!}{$
WD_p(\mathcal{PD}_k^1, \mathcal{PD}_k^2) = \min_{\phi: \mathcal{PD}_k^1 \rightarrow \mathcal{PD}_k^2} \left[ \sum_{x \in \mathcal{PD}_k^1} \|x - \phi(x)\|_{\infty}^{p}\right]^{\frac{1}{p}}
$}
\label{eq:WD}
\end{equation*}
where $\phi$ is the bijective mapping from $\mathcal{P}_k^1$ to $\mathcal{P}_k^2$ and $\| \cdot\|_\infty$ is the supremum norm. $WD_p$ is a form of optimal transport metric. For all the possible matching plans between  $\mathcal{P}_k^1$ and $\mathcal{P}_k^2$, $WD_p(\mathcal{P}_k^1, \mathcal{P}_k^2)$ is the infimum of the costs associated with matching plans~\cite{skraba2020wasserstein}. Hence, this metric quantifies the difference between the two \emph{PDs}, considering the pairing of points between off-diagonal points and diagonal points.

\begin{figure}[h!]
    \centering
    \includegraphics[width=0.9\linewidth]{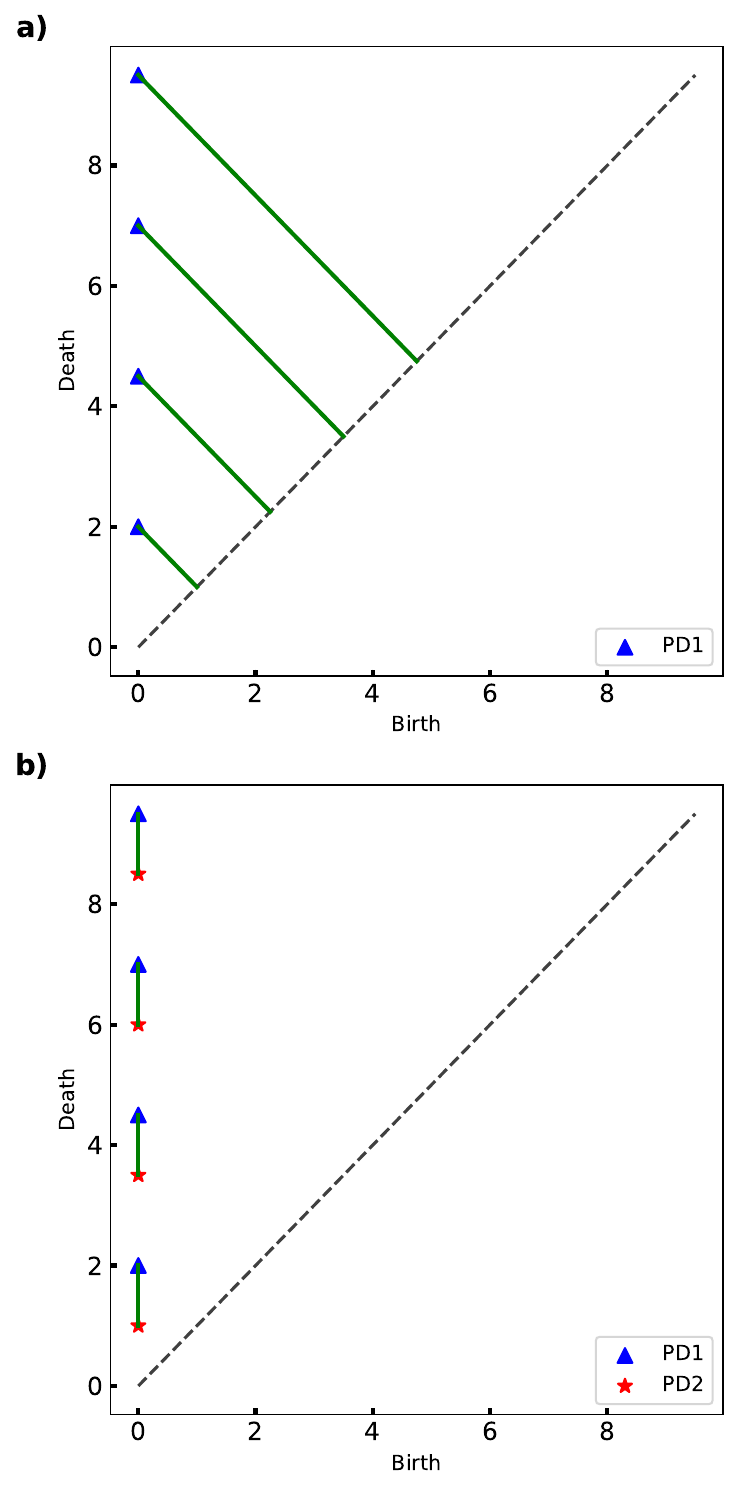}
    \caption{\justifying \small
    \textit{Figures depict the Wasserstein distance matching between persistence diagrams. The blue points correspond to PD1 and the red points correspond to PD2. Fig. (a) represents the comparison of PD1 points with the positive diagonal. Fig. (b) represents the comparison of PD1 with the PD2 points. Each solid green line represents the optimal matching for the corresponding point(s).}}
    \label{fig:WD}
\end{figure}

In this study, we use the $
WD_p$ metric in two different ways. Firstly, to study the evolution of topological dynamics of a market (or sector), we compare the \emph{PDs} corresponding to the sliding windows with the positive diagonal. This has been illustrated with an example in Fig.~\ref{fig:WD}(a). Fig.~\ref{fig:WD}(a) shows the matching between the points of a persistence diagram $PD_1$ with the positive diagonal. The green solid line represents the optimal matching distance between the points and the positive diagonal. The perpendicular distances between the points and the positive diagonal minimizes the cost of optimal transport. Secondly, we perform a direct comparison between the \emph{PDs} of two different markets or sectors corresponding to simultaneous sliding windows. This enables us to study the relative dynamics of two different markets or sectors at simultaneous periods of time. The same is illustrated with an example in Fig.~\ref{fig:WD}(b). Fig.~\ref{fig:WD}(b) shows the comparison between the points of two \emph{PDs} $PD_1$ and $PD_2$. Solid green lines connecting the points show the optimal mapping between the \emph{PDs}.

%The space of persistence diagrams can be endowed with a metric structure. One commonly used metric is the degree $p$-Wasserstein distance~\cite{gidea2018topological}, denoted $W_D(\mathcal{P}_k^1, \mathcal{P}_k^2)$, where $\mathcal{P}_k^1$ and $\mathcal{P}_k^2$ are two persistence diagrams at dimension $k$. The distance $W_D(\mathcal{P}_k^1, \mathcal{P}_k^2)$ is defined as: 
%\begin{equation}
%\label{$W_D$}
%W_D(\mathcal{P}_k^1, \mathcal{P}_k^2) = \min_{\phi: \mathcal{P}_k^1 \rightarrow \mathcal{P}_k^1} \left[ \sum_{x \in \mathcal{P}_k^1} \|x - \phi(x)\|_{\infty}^{p}\right]^{\frac{1}{p}}
%\end{equation}
%where $\phi$ is the bijective mapping from $\mathcal{P}_k^1$ to $\mathcal{P}_k^2$ and $\| \cdot\|_\infty $ is the sup norm. This metric measures the discrepancy between the two diagrams, taking into account the pairing of points between off-diagonal points and diagonal points. The Wasserstein distance provides a measure of similarity between persistence diagrams and is useful for comparing topological features across different datasets or analyzing the stability of features under perturbations.

%A key advantage of persistence homology is its robustness under small perturbations. If the underlying data changes slightly, the corresponding persistence diagram only moves a small Wasserstein distance from the diagram of the original data~\cite{gidea2018topological,cohen2005stability}. This property makes persistence homology a valuable tool for analyzing complex systems.

\subsection{Granger-Causality}
\label{Sec:GC}
The Granger-causality test is a statistical method used to assess whether one time series provides valuable information for forecasting another. It is widely applied in econometrics and financial studies to infer directional relationships between variables. The test relies on the assumption that if a variable \( X_t \) Granger-causes \( Y_t \), then the past values of \( X_t \) should contain information that helps predict \( Y_t \) beyond the information contained in the past values of \( Y_t \) alone \cite{Granger1969investigating,shojaie2022Granger,barrett2010multivariate}.

Let $X_t$ and  $Y_t$ be two stationary time-series. The Granger causality test involves estimating the following vector autoregressive (VAR) models~\cite{kirchgassner2012introduction}:
\begin{align}
    Y_t &= \alpha_0 + \sum_{i=1}^p \alpha_i Y_{t-i} + \epsilon_t, \label{eq:univariate_model} \\
    Y_t &= \beta_0 + \sum_{i=1}^p \beta_i Y_{t-i} + \sum_{i=1}^q \gamma_i X_{t-i} + \eta_t, \label{eq:bivariate_model}
\end{align}
where $\epsilon_t$  and $\eta_t$ are white noise error terms,  $p$ and $q$ denote the lag orders, and $ \alpha_i$, $\beta_i$, and $\gamma_i$ are coefficients.

Eq.~\ref{eq:univariate_model} corresponds to a univariate model while Eq.~\ref{eq:bivariate_model} corresponds to a bivariate model. This means that the value of $Y_t$ is predicted using the past values of $Y$ up to a lag of $p$ in the univariate model while the bivariate model incorporates the past values of $X$ in addition to that of $Y$ to determine $Y_t$. The improvement in the prediction of $Y_t$ with model \ref{eq:bivariate_model} as compared to model \ref{eq:univariate_model} indicates the influence of $X_t$ on $Y_t$~\cite{smirnov2009granger}. \\
To test whether  $X_t$ Granger-causes $Y_t$, we perform an $F$-test on the null hypothesis:
\begin{equation}
    H_0: \gamma_i = 0 \quad \forall i \in \{1, 2, \dots, q\}.
\end{equation}
Here, $H_0$  states that $X_t$ does not Granger-cause $Y_t$, meaning that the past values of $X_t$ provide no additional predictive information about $Y_t$ beyond the past values of $Y_t$ itself. The null hypothesis is rejected if the p-value is below a significance level~\cite{kirchgassner2012introduction}. 

The lag orders $p$ and $q$ are critical to the analysis. We take the same value for $p$ and $q$ in this analysis. Values are typically chosen using the Final Prediction Error (FPE) method \cite{lutkepohl2005new,amiri2016income}. The FPE method, minimizes the prediction error of the VAR model, ensuring optimal lag selection~\cite{thornton1985lag,liew2004lag}. Ensuring the stationarity of the time-series is also crucial, which can be verified using the Augmented Dickey-Fuller (ADF) test \cite{dickey1979distribution} or the Phillips-Perron test \cite{phillips1988testing}. If the time-series are non-stationary, differencing techniques may be applied to achieve stationarity \cite{engle1987co,kang2013linkage}.

%Granger causality has found extensive application in financial markets to study interactions between Stock, commodity, currency, and interest rate indices \cite{bessler2003price, chen2009global, hammoudeh2015relationships}.
%\section{TDA Approach to Analyzing Multiple time-series}
%\label{our Approach}

\section{Data Analyzed}
\label{Data}
We took three years (01-06-2018 to 01-06-2021) of data from twenty US stocks with the highest market capitalization from all major sectors and twenty commodity constituents of S\&P GSCI Index (\^{}SPGSCI). The names and tickers of stocks and commodities are listed in Table \ref{tab:stocks_commodities}. We also conducted a sectoral analysis of the US Stock market for eleven sectors based on The Global Industry Classification Standard (GICS). We selected the five highest market-capitalized stocks of each sector. The names and tickers of the stocks taken from each sector are listed in Table. \ref{tab:us_sectors_stocks}. We use the python package \emph{Giotto-TDA}~\cite{tauzin2021giotto} for topological computations. The dates in Figs.~\ref{fig:WD_Stock}-\ref{wd_comp_sectors} are in (mm-yyyy) format.\\

\section{Granger-causal Analysis using TDA}
\label{gc_appraoch}
In this section, we briefly explain the steps taken to perform the Granger-causal analysis of financial markets and sectors using TDA. The sequence of steps taken to identify crashes in financial markets using TDA by considering multiple time-series was shown in Ref.~\citenum{rai2024identifying}.\\

Initially, we construct an $n$-dimensional Euclidean point cloud by calculating the log-return of $n$ stock (or commodity) price time-series. The log-return of $n$ stock price time-series of length $l$ is calculated as $R_{i,j}\coloneqq\ln{\frac{P_{i,j}}{P_{i,j-1}}}$, where $P_{m,k}$ denotes the closing price of the index $m=\{1,\ldots,n\}$ on day $k \in \{1,\ldots,l\}$. By taking the log-return of all time-series on the $j$-th day, i.e., $(R_{1,j}, R_{2,j}, \ldots,R_{n,j})$, a point is formed in $\mathbb{R}^n$. An $n$-dimensional Euclidean point cloud is obtained by taking all such points for $j=\{1,2,\ldots,l\}$:

$$D=\left\{\left(X_{1,j}, X_{2,j}, \ldots, X_{n,j}\right)\right\}_{j=1}^l \subset \mathbb{R}^n.$$

A sliding window of size $w$ is applied to subsets of this point cloud, and for each window position, the Rips filtration is constructed across various scales ($\varepsilon$). The birth and death of $0$-dimensional holes are recorded in persistence diagrams (\emph{PDs}) for each window position, resulting in $(l-w)$ many persistence diagrams.

These \emph{PDs} are analyzed using the Wasserstein Distance ($WD_p$) metric. The $WD_p$ between the persistence diagram for each window with the positive diagonal is calculated. This distance measures the topological similarity between two PDs, and its evolution helps in understanding the dynamics of multiple time-series. The $WD_p$ are calculated for each persistence diagram, and their variation over time is studied to comprehend the dynamics of the multiple time-series.

We analyze the Granger-causal relationships between the time-series of $WD_p$ of Stock and commodity markets. A similar analysis is conducted for the different sectors of the US stock market. This approach provides a comprehensive view of the influence between the different financial markets and sectors.

The choice of the window size ($w$) is usually dictated by the time scale of the data and the application at hand. In the study of stock-market dynamics, common window sizes are 20, 30, and 60 days~\cite{guo2020empirical,goel2020topological}. In our analyses, we chose a window size of 30 days. We also tried 45 and 60 days, but the outcomes were not very sensitive to these alternative choices. For best results, we used 30 days throughout the paper.

\section{Results and Discussion}
\label{result}

This section contains the results of the identification of crashes in the stock and commodity markets during the COVID-19 pandemic using TDA. TDA facilitates the use of multidimensional time-series at once and hence allows the identification of crashes for the entire market by considering multiple stocks/commodities. Also, we perform a direct topological comparison between the Stock and commodity markets. Subsequently, we categorize the time-series obtained from TDA into three periods viz., pre-crash, crash, and post-crash, each spanning a year. For each period we test the Granger-causal relationships between Stock and commodity markets to determine the direction of causality, if any, and identify which market influences the other. We extend our study to identify the market crashes across different US sectors. Further, we perform a direct topological comparison between the sectors. Finally, we test the Granger-causal relationships between different sectors to analyze which sectors were more influenced or sensitive during different periods.

\subsection{Identification of crashes}
This section identifies the crashes in different markets and sectors during the COVID-19 pandemic using TDA.

\label{Identification_crashes}
\subsubsection{Stock market}
\label{Stock_Identification}
\begin{figure}[ht]
    \centering
    \includegraphics[width=0.9\linewidth]{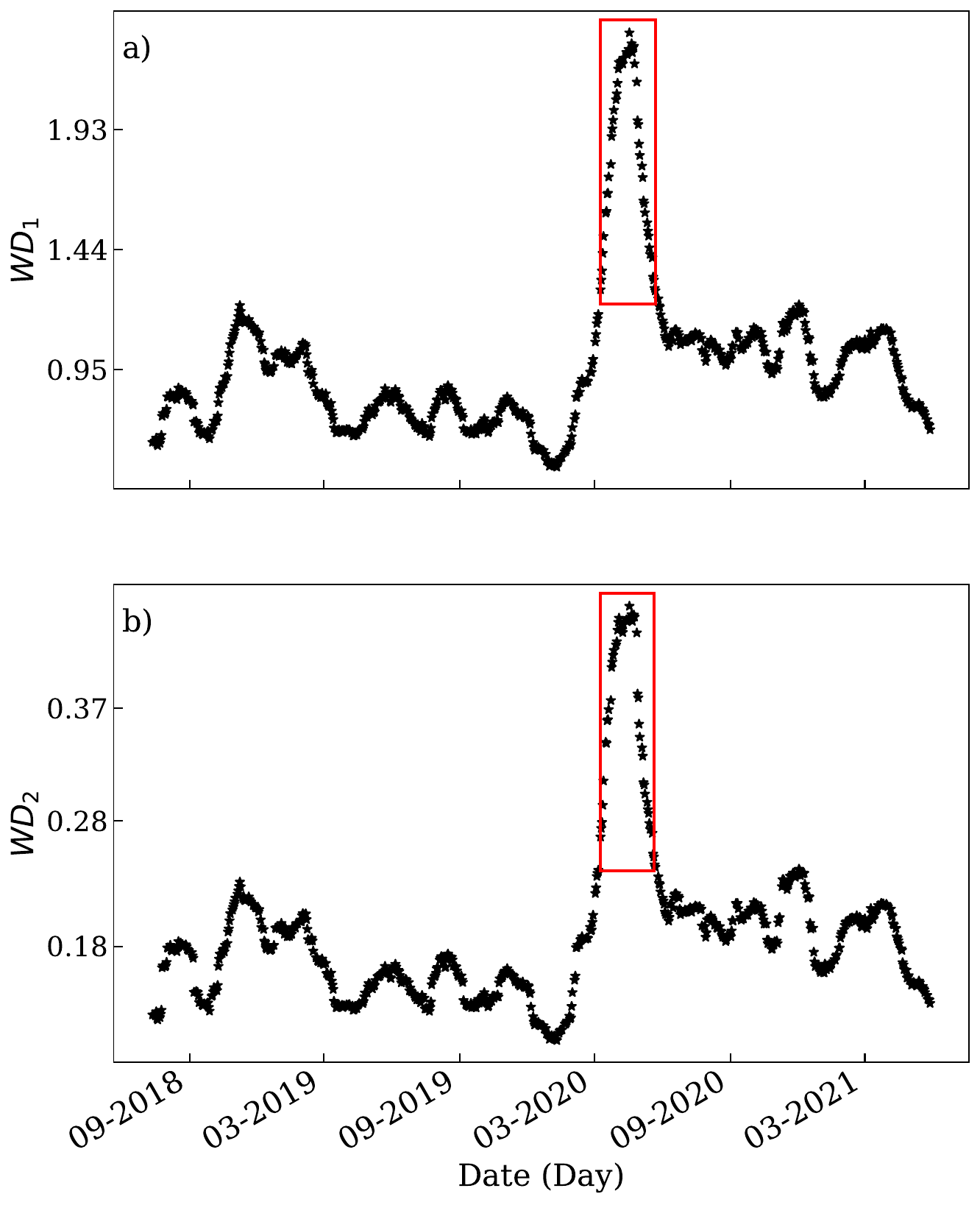}
    \caption{\justifying \small
    \textit{Plot (a) represents the \textit{degree-1} Wasserstein distance and plot (b) represents \textit{degree-2} Wasserstein distances in the Stock market. Both plots display abrupt spikes during the COVID-19 pandemic indicated by a red box.}}
    \label{fig:WD_Stock}
\end{figure}

Figs.~\ref{fig:WD_Stock}(a) and \ref{fig:WD_Stock}(b) represent the \textit{degree-1} Wasserstein distance ($WD_1$) and \textit{degree-2} Wasserstein distance ($WD_2$) of the stock market. These metrics, obtained from TDA are applied to capture topological structural changes in the underlying data. A significant rise in the \emph{WDs'} is observed during early 2020 which corresponds with the onset of the COVID-19 pandemic. The pandemic has led to uncertainty and volatility in the stock market due to strict lockdown restrictions. The abrupt rise in the \emph{WDs'} signifies a drastic change in the topological structure in the Stock market indicating a change of pattern in the market behavior. These findings show that a major crash occurred in the Stock market during the COVID-19 pandemic.

%TDA confirms that a crash occured in the Stock market due to the COVID-19 pandemic.

\subsubsection{Commodity market}
\label{Commodity_Identification}
\begin{figure}[h]
    \centering
    \includegraphics[width=0.9\linewidth]{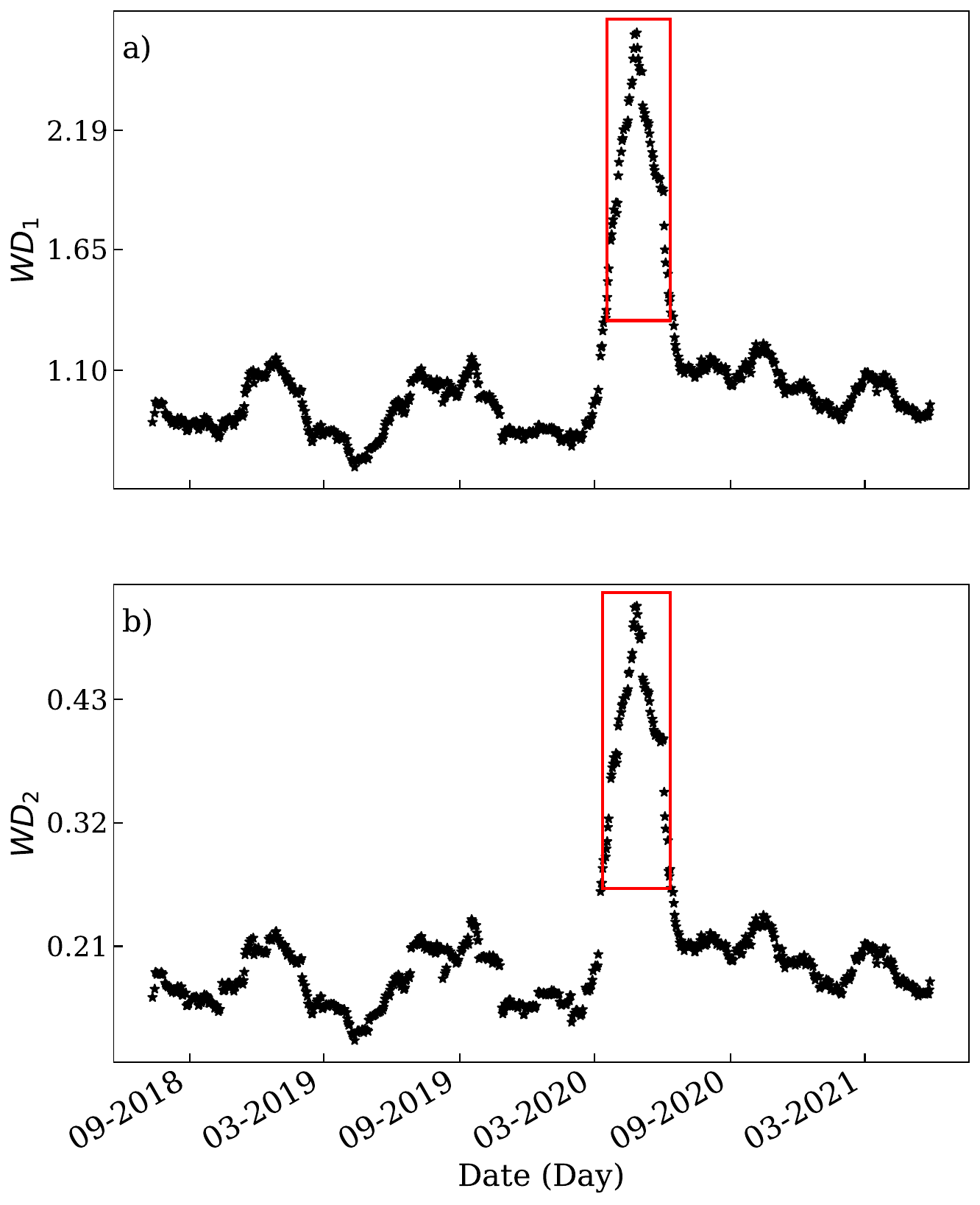}
    \caption{\justifying \small  \textit{
    Plot (a) represents the degree-1 Wasserstein distance and plot (b) represents \textit{degree-2} Wasserstein distances in the commodity market. Both plots display abrupt spikes during the COVID-19 pandemic indicated by a red box.}}
    \label{fig:Comm_Stock}
\end{figure}

Figs.~\ref{fig:Comm_Stock}(a) and \ref{fig:Comm_Stock}(b) represent the \textit{degree-1} Wasserstein distance ($WD_1$) and \textit{degree-2} Wasserstein distance ($WD_2$) of the commodity market. The topological structure of the commodity market during the COVID-19 pandemic changed significantly when compared with the topology during the normal period which is quantified by the \emph{WD}. A clear steep rise in the \emph{WDs'} is observed during early 2020 which aligns with the onset of the COVID-19 pandemic. This proves that the COVID-19 pandemic led to a huge crash in the commodity market.

\subsubsection{Sectors}
\label{Sectors_identification}
\begin{figure}[H]
    \centering
    \includegraphics[width=0.9\linewidth]{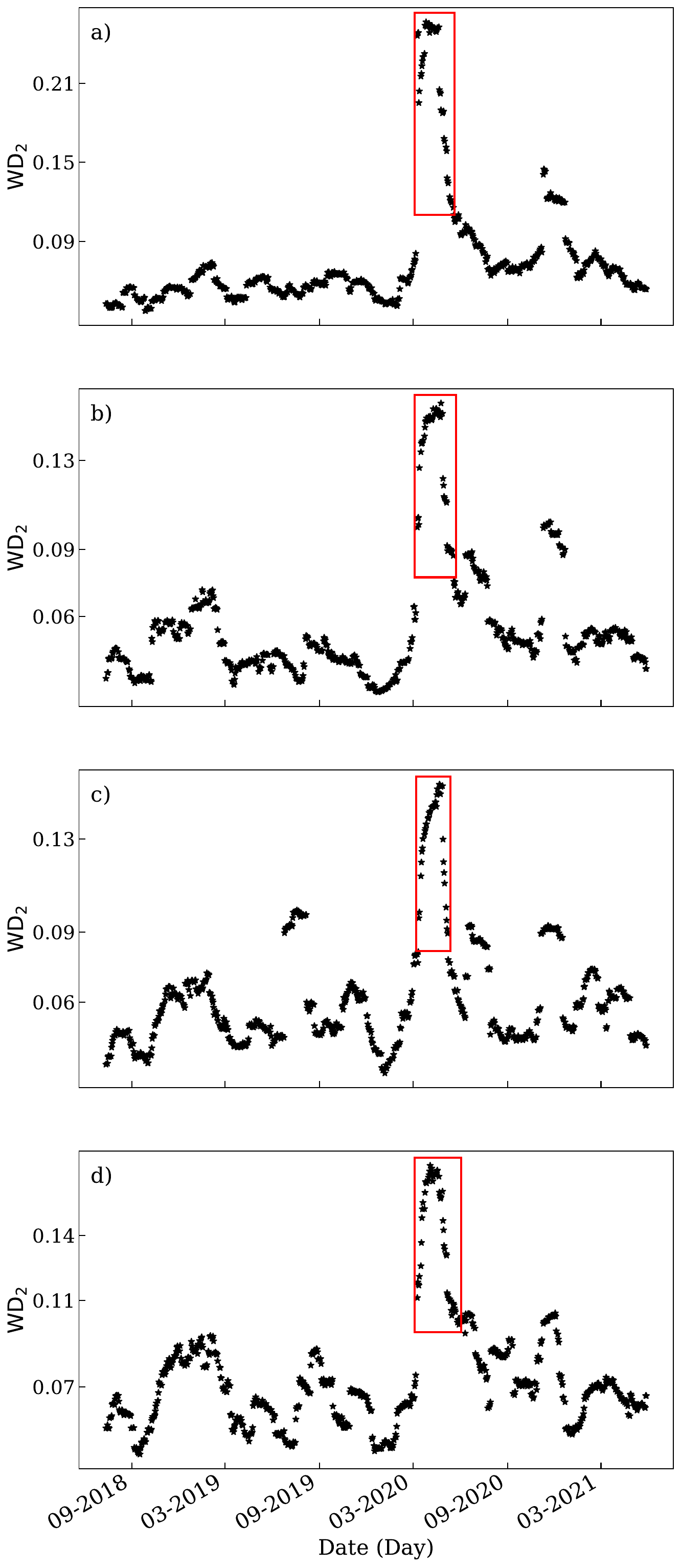}
    \caption{\justifying \small \textit{ Degree-2 Wasserstein Distances for different sectors. Plot (a) corresponds to the Energy sector. Plot (b) corresponds to the  Financial sector. Plot (c) corresponds to the Healthcare sector and Plot (d) corresponds to the industrial sector. All the plots show an abrupt spike during the market crash due to COVID-19 indicated by a red box. }}
    \label{fig:Sectors_$WD_2$}
\end{figure}

We have identified crashes in different US sectors due to the COVID-19 pandemic. Major sectors like Consumer Discretionary, Consumer Staples, Energy, Financials, Healthcare, Industrials, IT, Materials, Real Estate, Utilities, and Communication services are considered for the analysis. 

For each sector, we have considered five stocks based on market capitalization. The time-series of these five stocks are used to construct a point-clould in 5-dimensional space (mathematical). For a sliding window of 30 days, persistence diagrams (\emph{PDs}) are constructed from the point-cloud dataset and Wasserstein Distance (\emph{WD}) is calculated by comparing each PD with the positive diagonal.

Figs.~\ref{fig:Sectors_$WD_2$}a-~\ref{fig:Sectors_$WD_2$}d represents the {degree-2} \emph{WDs'} for the Energy, Finance, Healthcare and Industrial sectors, respectively. Clear spikes were observed in all sectors during early 2020 which corresponds with the onset of the COVID-19 pandemic. This shows that these sectors were adversely impacted during the COVID-19 pandemic. Similar results are observed for the rest of the sectors. We observe that the fluctuation in \emph{WD} is higher in individual sectors than in overall Stock or commodity markets. This clearly shows that sectors exhibit higher volatility than the broader markets.

\subsection{Topological comparison}
\label{topology_comp}
In Sec.~\ref{Identification_crashes}, we identified crashes due to the COVID-19 pandemic in stock, commodity, and different US sectors by comparing their respective topological structure with the trivial PD i.e. the positive diagonal. However, TDA also offers the scope of directly comparing the topologies of two different markets (or sectors) using the \emph{Wasserstein Distance (WD)} metric, as explained in Sec.~\ref{WassersteinD}. A small \emph{WD} indicates strong topological similarity, whereas, a spike reveals significant topological disparity.

\subsubsection{Stock vs Commodity}
\label{comp_Stock_commodity}
%We have already detected crash due to COVID-19 pandemic in Stock and commodity markets, in Sec.~\ref{Stock_Identification} and ~\ref{Commodity_Identification}, by comparing their respective topological structure with a reference structure during normal period. The analysis demonstrated that the topology were significantly different, indicating a market crash.  we compare the topological structures of Stock and commodity markets with each other. One of the objectives of the comparision is to check whether there is any temporal lag in the change of topology between the two markets. Second objective of the comparision is to analyze the behavior of the two markets and check whether they experienced crashes of similar intensity. If the topology is similar and there is no temporal lag in both markets, we would get minimal fluctuations in the WD, however, if the structures are different we will see huge fluctuations in the WD. The presence of huge fluctuations hints towards the flow of information in different markets.

 %The analysis demonstrated that the topology were significantly different, indicating a market crash. The difference was quantified by WD. TDA offers the scope of directly comparing the topology of Stock and commodity market by calculating the Wasserstein Distance between the persistence diagrams. We calculate the \textit{degree-2} Wasserstein distance ($WD_2$) between the 0-dimensional persistence diagrams of Stock and Commodity market with a sliding window of 30 days. 
This section compares the topology of the stock and commodity markets from June 2018 to June 2021. We construct the $0$-dimensional persistence diagrams (\emph{PDs}) for the stock and commodity markets separately with a sliding window of $30$ days using similar steps as mentioned in Sec.~\ref{Identification_crashes}. Subsequently, we calculate the \emph{degree-2 Wasserstein Distance $WD_2$} between the $0$-dimensional persistence diagrams of stock and commodity market for the corresponding time periods.  
\begin{figure}[h]
    \centering
    \includegraphics[width=0.9\linewidth]{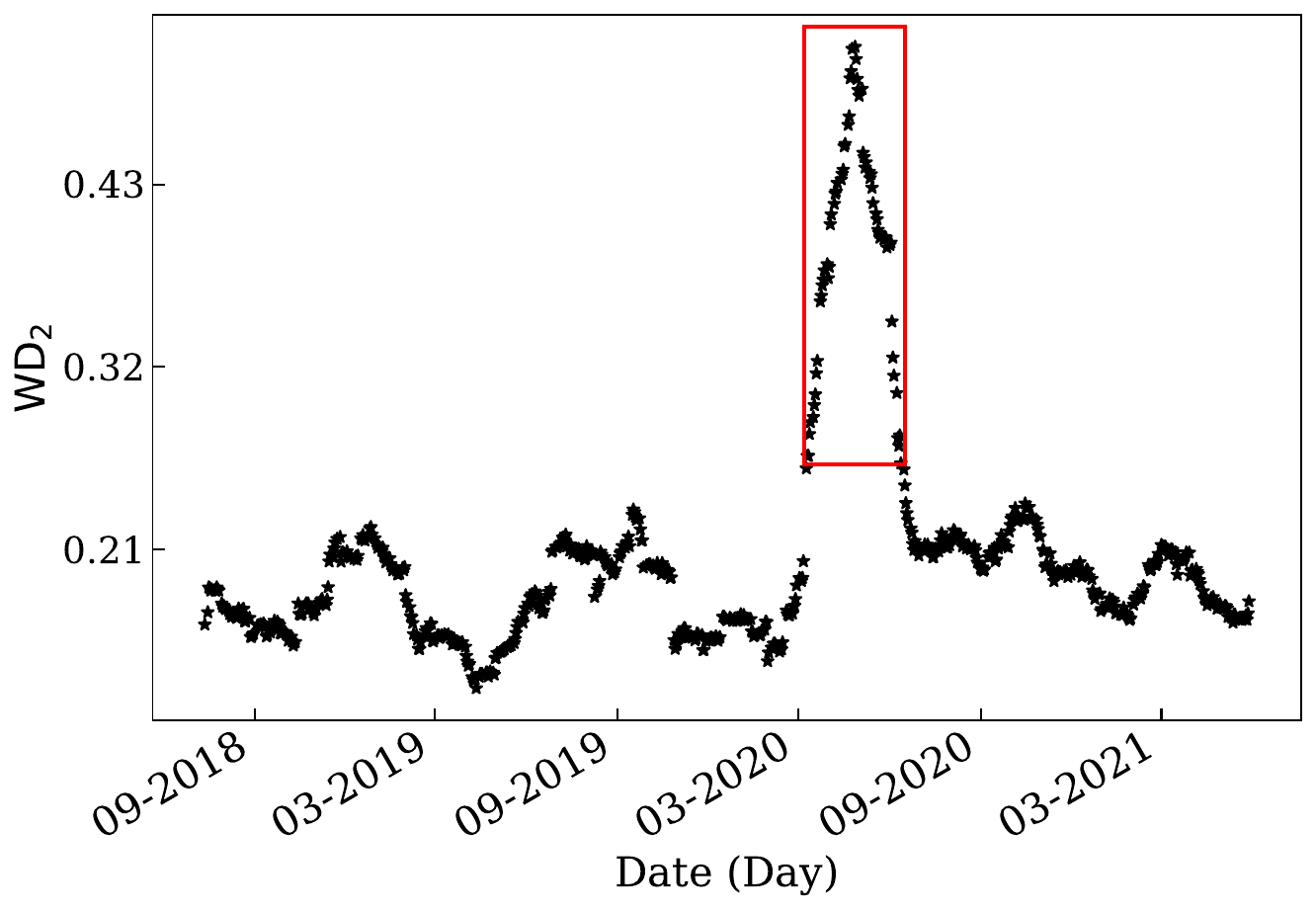}  % Adjust the caption settings
    \caption{\justifying \small
    \textit{Plot represents the $WD_2$ distance between the corresponding 0-dimensional persistence diagrams of stock and commodity markets from June 2018 to June 2021. A significant spike is observed in the $WD_2$ during the crash period indicated by a red box.}} 
    \label{fig:$WD_2$_Stock_comm}
\end{figure}

Fig.~\ref{fig:$WD_2$_Stock_comm} shows the \emph{$WD_2$} plot between the corresponding $0$-dimensional persistence diagrams of stock and commodity markets from June 2018 to June 2021. We observe a significant spike in the $WD_2$ during the crash period due to COVID-19. This signifies that during the crash period, there was a marked topological difference between the stock and commodity markets. The significant topological difference between the two markets may arise due to the difference in the topological magnitude of the crash; see Sec.~\ref{topological_magnitude}. This may also occur due to the temporal lag in the crashes suggesting \emph{information flow} between the two markets~\cite{shahzad2021impact}, which Sec.~\ref{gc} studies in detail using Granger-causality.

\subsubsection{Sectors vs Sectors}
\label{comp_sectors}
\begin{figure}[h]
    \centering
    \includegraphics[width=0.9\linewidth]{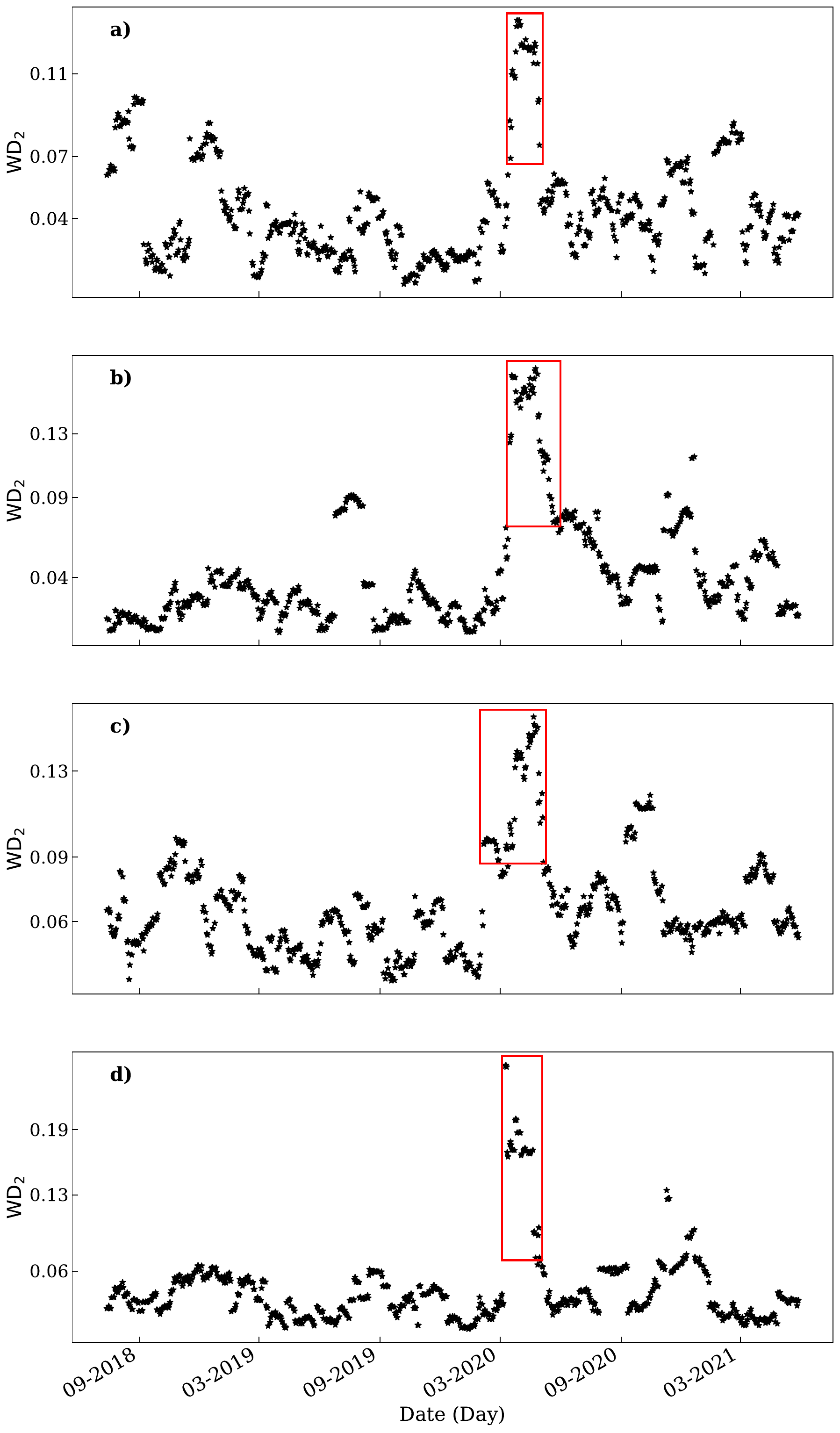}
    \caption{\justifying \small
    \textit{Degree-2 Wasserstein distances between the $0$-dimensional persistence diagrams of different sectors. Fig. (a) represents the $WD_2$ between the  Communication Services and IT sectors. Fig. (b) represents the $WD_2$ between Real Estate and Healthcare sectors. Fig. (c) represents the $WD_2$ between Consumer Staples and Consumer Discretionary sectors. Fig. (d) represents the $WD_2$ between Industrials and Energy sectors. A significant spike is observed in all the figures during the market crash induced by COVID-19.}}
    \label{wd_comp_sectors}
\end{figure}
The topology of each sector is also compared which showed clear spikes during the COVID-19 pandemic. The significant difference in the topology between sectors reveals the presence of causation and influences the nature of some sectors to others\cite{adekoya2021covid}. Another reason may be the presence of some sensitive sectors that move due to the influence of other sectors\cite{}. This leads to a lag in the movement resulting in such differences in topology. Hence, it becomes necessary to check for interdependence and their causal relationships.

Figs.~\ref{wd_comp_sectors}(a)-~\ref{wd_comp_sectors}(d) illustrates the pairwise relative topological changes between different sectors of the US stock market. Fig.~\ref{wd_comp_sectors}(a) represents the $WD_2$ between the  Communication Services and IT sectors. Fig.~\ref{wd_comp_sectors}(b) represents the $WD_2$ between Real Estate and Healthcare sectors. Fig.~\ref{wd_comp_sectors}(c) represents the $WD_2$ between Consumer Staples and Consumer Discretionary sectors. Fig.~\ref{wd_comp_sectors}(d) represents the $WD_2$ between Industrials and Energy sectors. Topological differences are determined using the \textit{degree-2} Wasserstein distance ($WD_2$) between the simultaneous persistence diagrams of the 0-dimensional homology group for each sector. We observe a significant spike in each plot, which highlights how the topology of a given sector during the market crash is markedly different from the topology of another sector. Although all sectors are affected by the crash, the corresponding topology during the crash period remains significantly different for different sectors. Moreover, the direct comparison of topology between sectors shows greater fluctuations compared to the mapping with the positive diagonal as shown in Fig.~\ref{fig:Sectors_$WD_2$}. Hence, the dynamic relation between sectors fluctuates greatly compared to the fluctuations in the dynamics of a given sector.

\subsection{Topological magnitudes of the crash}
\label{topological_magnitude}
In Sec.~\ref{topology_comp}, we saw that the direct topological comparison of the two markets (or sectors) produced significant spikes during the crash period. One reason for the spike could be the different magnitudes of the crash. In this section, we compare the topological magnitudes of the crash between different markets (or sectors) by comparing the mean and maximum values of the \emph{Wasserstein Distance} metric corresponding to the respective markets and sectors.

\subsubsection{Stock and Commodity}
\label{topological_magnitude_Stock_commmodity}
As the \emph{Wasserstein Distance} between the stock and commodity markets showed significant spikes during the crash period, we compare their respective $WD_1$ and $WD_2$ time-series as shown in Table~\ref{tab:mean_max_WD2}.\\
%\textcolor{red}{think about normalizing the values.}
\begin{table}[h!]
\centering
\resizebox{0.3\textwidth}{!}{%
\begin{tabular}{lcccc}
\hline \hline
 & \multicolumn{2}{c}{\textbf{Stock}} & \multicolumn{2}{c}{\textbf{Commodity}} \\ \hline
Metric & $WD_1$ & $WD_2$ & $WD_1$ & $WD_2$ \\ \hline
Mean & \multicolumn{1}{c}{1.00} & \multicolumn{1}{c}{1.00} & \multicolumn{1}{c}{1.08} & \multicolumn{1}{c}{1.05} \\
Max & \multicolumn{1}{c}{2.43} & \multicolumn{1}{c}{2.37} & \multicolumn{1}{c}{2.74} & \multicolumn{1}{c}{2.68} \\
\hline \hline
\end{tabular}%
}
\caption{\justifying \small
\textit{Table shows the mean and maximum values of the $WD_1$ and $WD_2$ time-series corresponding to Stock and commodity markets. The values for both metrics are normalized to their mean values for the stock market.}}
\label{tab:mean_max_WD2}
\end{table}

Table~\ref{tab:mean_max_WD2} shows the mean and maximum of the \emph{WD} time-series corresponding to stock and commodity markets normalized to their means for the stock market. We observe that the mean of the corresponding  $WD_1$ and $WD_2$ time-series are comparable for the two markets. Also, the maximum value of these metrics---during the crash period---is insignificantly larger for the commodity market as compared to the stock market.\\

\subsubsection{Sectors}
\label{topological_magnitude_sectors}
The \emph{Wasserstein Distance} between US sectors showed significant spikes during the crash period. Therefore, we compare the magnitudes of the crash between different sectors by comparing the mean and maximum values of the $WD_2$ time-series. The values are listed in Table~\ref{tab:mean_max_WD2}.

\begin{table}[H]
\centering
\resizebox{0.4\textwidth}{!}{%
\begin{tabular}{lcc}
\hline \hline
\textbf{Sector} & \textbf{Mean $WD_2$} & \textbf{Max $WD_2$} \\
\hline
Industrials & \multicolumn{1}{c}{0.07} & \multicolumn{1}{c}{0.17} \\
Energy & \multicolumn{1}{c}{0.07} & \multicolumn{1}{c}{0.26} \\
Financial & \multicolumn{1}{c}{0.05} & \multicolumn{1}{c}{0.16} \\
Healthcare & \multicolumn{1}{c}{0.06} & \multicolumn{1}{c}{0.15} \\
Consumer Discretionary & \multicolumn{1}{c}{0.08} & \multicolumn{1}{c}{0.21} \\
Consumer Staples & \multicolumn{1}{c}{0.05} & \multicolumn{1}{c}{0.13} \\
Real Estate & \multicolumn{1}{c}{0.07} & \multicolumn{1}{c}{0.24} \\
Utilities & \multicolumn{1}{c}{0.05} & \multicolumn{1}{c}{0.16} \\
Materials & \multicolumn{1}{c}{0.06} & \multicolumn{1}{c}{0.19} \\
Information Technology & \multicolumn{1}{c}{0.07} & \multicolumn{1}{c}{0.20} \\
Communications Services & \multicolumn{1}{c}{0.08} & \multicolumn{1}{c}{0.15} \\
\hline \hline
\end{tabular}%
}
\caption{\justifying \small
\textit{Table shows the mean and maximum values of $WD_2$ time-series for all Sectors.}}
\label{tab:mean_max_WD2}
\end{table}

Table~\ref{tab:mean_max_WD2} shows the mean and maximum values of $WD_2$ time-series for different sectors. The mean values for all sectors are in the range of 0.05 to 0.08. However, the maximum values corresponding to the crash period are significantly different for different sectors ranging from 0.13 to 0.26. This means that the different sectors had different magnitudes of the topological crash.

\subsection{Granger-Causality}
\label{gc}
In Sec.~\ref{topology_comp}, we saw that the relative topological difference between stock and commodity markets is significant during the crash period. Similar significant spikes are observed between different US sectors during the crash period. One of the possible reasons for the spikes could be the temporal lag in the topological dynamics of the different markets (or sectors). This could indicate a possible dominance of one market (or sector) over the other for a given period of time. Also, it could be because of the sensitive nature of some sectors due to the negative impact induced by the COVID-19 pandemic. Analyzing the causality during different periods of market crash is important as it could indicate the information flow in the market during the crisis period. Moreover, the TDA-based study incorporates multiple time-series of a given market (or sector) and hence presents a complete analysis. We study Granger-causality for three different time periods viz. pre-crash (01-06-2018 to 01-06-2019), crash (01-06-2019 to 01-06-2020), and post-crash (01-06-2020 to 01-06-2021).\\
%\textcolor{red}{Is it the assumption before Granger or the assumption of Granger causality that the time-series should be stationary. And for that we make the time-series stationary}
The use of Granger-causality is appropriate when the time-series is stationary~\cite{shojaie2022Granger}. The Augmented Dickey-Fuller (ADF)  and Phillips Perron (PP) tests are common statistical tests used to check for stationarity in a time-series~\cite{kang2013linkage}. The test uses the existence of a unit root as the null hypothesis. The test result is analyzed using their respective test statistic and p-values. In this study, we reject the null hypothesis at a $5\%$ significance level meaning that when the p-values for the respective tests are less than 0.05, we reject the non-stationarity of time-series and accept the time-series as stationary. If times series are non-stationary with unit roots, they must be made stationary by means of a difference filter before carrying out Granger causality tests~\cite{kang2013linkage}.
%\textcolor{red}{mention which type: zero-mean, single-mean, or linear-time trend.}

\subsubsection{Stationarity Tests - Stock and Commodity}
\label{stationarity_Stock_commodity}

We first perform the ADF and PP test to check whether the time-series of Wasserstein distances $WD_1$ and $WD_2$ for Stock and commodity markets are stationary or not. The results of the ADF and PP tests for both stock and commodity markets are presented in Table~\ref{tab:stationarity_analysis}.

\begin{table}[h!]
\centering
\resizebox{0.5\textwidth}{!}{%
\begin{tabular}{lrrrrrr}
\hline \hline
\textbf{Metric} & \textbf{Sector} & \textbf{ADF Stat} & \textbf{ADF p-val} & \textbf{PP Stat} & \textbf{PP p-val} \\
\hline
\multicolumn{1}{l}{\textbf{Pre-crash period}} & \multicolumn{1}{c}{} & \multicolumn{1}{c}{} & \multicolumn{1}{c}{} \\
\hline
$WD_1$ & Commodity & \multicolumn{1}{c}{-1.51} & \multicolumn{1}{c}{0.53} & \multicolumn{1}{c}{-1.57} & \multicolumn{1}{c}{0.50} \\
$WD_1$ & Stock     & \multicolumn{1}{c}{-1.57} & \multicolumn{1}{c}{0.50} & \multicolumn{1}{c}{-1.75} & \multicolumn{1}{c}{0.40} \\
$WD_2$ & Commodity & \multicolumn{1}{c}{-1.28} & \multicolumn{1}{c}{0.64} & \multicolumn{1}{c}{-1.59} & \multicolumn{1}{c}{0.49} \\
$WD_2$ & Stock     & \multicolumn{1}{c}{-1.66} & \multicolumn{1}{c}{0.45} & \multicolumn{1}{c}{-1.87} & \multicolumn{1}{c}{0.35} \\
\hline
\multicolumn{1}{l}{\textbf{Crash period}} & \multicolumn{1}{c}{} & \multicolumn{1}{c}{} & \multicolumn{1}{c}{} \\
\hline
$WD_1$ & Commodity & \multicolumn{1}{c}{-2.09} & \multicolumn{1}{c}{0.25} & \multicolumn{1}{c}{-0.61} & \multicolumn{1}{c}{0.87} \\
$WD_1$ & Stock     & \multicolumn{1}{c}{-1.89} & \multicolumn{1}{c}{0.34} & \multicolumn{1}{c}{-1.35} & \multicolumn{1}{c}{0.60} \\
$WD_2$ & Commodity & \multicolumn{1}{c}{-1.54} & \multicolumn{1}{c}{0.51} & \multicolumn{1}{c}{-0.53} & \multicolumn{1}{c}{0.89} \\
$WD_2$ & Stock     & \multicolumn{1}{c}{-1.70} & \multicolumn{1}{c}{0.43} & \multicolumn{1}{c}{-1.38} & \multicolumn{1}{c}{0.59} \\
\hline
\multicolumn{1}{l}{\textbf{Post-crash period}} & \multicolumn{1}{c}{} & \multicolumn{1}{c}{} & \multicolumn{1}{c}{} \\
\hline
$WD_1$ & Commodity & \multicolumn{1}{c}{-1.31} & \multicolumn{1}{c}{0.62} & \multicolumn{1}{c}{-1.62} & \multicolumn{1}{c}{0.47} \\
$WD_1$ & Stock     & \multicolumn{1}{c}{-2.91} & \multicolumn{1}{c}{0.04**} & \multicolumn{1}{c}{-1.85} & \multicolumn{1}{c}{0.36} \\
$WD_2$ & Commodity & \multicolumn{1}{c}{-2.73} & \multicolumn{1}{c}{0.07} & \multicolumn{1}{c}{-1.61} & \multicolumn{1}{c}{0.48} \\
$WD_2$ & Stock     & \multicolumn{1}{c}{-3.11} & \multicolumn{1}{c}{0.03**} & \multicolumn{1}{c}{-2.03} & \multicolumn{1}{c}{0.27} \\
\hline \hline
\end{tabular}%
}
\caption{\justifying \small
\textit{Stationarity Test Results for $WD_1$ and $WD_2$ time-series across different periods for Stock and commodity market. The table shows statistics and p-values for the ADF and PP stationarity test. At a $5\%$ significance level, the time-series are not consistently stationary for both tests and across all periods.}}
\label{tab:stationarity_analysis}
\begin{flushleft}
\textit{** indicates significance at the $5\%$ significance level.}
\end{flushleft}
\end{table}

%\textcolor{red}{For stationary time-series what are the condition and values write about P-val or other}

Table \ref{tab:stationarity_analysis} shows that the p-values for the time-series $WD_1$ and $WD_2$ are not consistently less than 0.05 for both markets over the three time periods. Therefore, we cannot reject the null hypothesis, and hence both the time-series are non-stationary. As a result, we tested the first difference time-series of $WD_1$ and $WD_2$ for stationarity using the ADF and PP test. The results are listed in the Table~\ref{tab:stationarity_analysis_fd}.
\begin{table}[h!]
\centering
\resizebox{0.5\textwidth}{!}{%
\begin{tabular}{lrrrrrr}
\hline
\hline
\textbf{Metric} & \textbf{Sector} & \textbf{ADF Stat} & \textbf{ADF p-val} & \textbf{PP Stat} & \textbf{PP p-val} \\
\hline
\multicolumn{1}{l}{\textbf{Pr-crash period}} & \multicolumn{1}{c}{} & \multicolumn{1}{c}{} & \multicolumn{1}{c}{} \\
\hline
$WD_1$ & Commodity & \multicolumn{1}{c}{-15.42} & \multicolumn{1}{c}{0.00**} & \multicolumn{1}{c}{-15.52} & \multicolumn{1}{c}{0.00**} \\
$WD_1$ & Stock     & \multicolumn{1}{c}{-11.10} & \multicolumn{1}{c}{0.00**} & \multicolumn{1}{c}{-11.99} & \multicolumn{1}{c}{0.00**} \\
$WD_2$ & Commodity & \multicolumn{1}{c}{-15.34} & \multicolumn{1}{c}{0.00**} & \multicolumn{1}{c}{-15.49} & \multicolumn{1}{c}{0.00**} \\
$WD_2$ & Stock     & \multicolumn{1}{c}{-5.96}  & \multicolumn{1}{c}{0.00**} & \multicolumn{1}{c}{-12.94} & \multicolumn{1}{c}{0.00**} \\
\hline
\multicolumn{1}{l}{\textbf{Crash period}} & \multicolumn{1}{c}{} & \multicolumn{1}{c}{} & \multicolumn{1}{c}{} \\
\hline
$WD_1$ & Commodity & \multicolumn{1}{c}{-3.64}  & \multicolumn{1}{c}{0.01**} & \multicolumn{1}{c}{-12.64} & \multicolumn{1}{c}{0.00**} \\
$WD_1$ & Stock     & \multicolumn{1}{c}{-3.06}  & \multicolumn{1}{c}{0.03**} & \multicolumn{1}{c}{-9.58}  & \multicolumn{1}{c}{0.00**} \\
$WD_2$ & Commodity & \multicolumn{1}{c}{-2.40}  & \multicolumn{1}{c}{0.14}   & \multicolumn{1}{c}{-14.02} & \multicolumn{1}{c}{0.00**} \\
$WD_2$ & Stock     & \multicolumn{1}{c}{4.29}   & \multicolumn{1}{c}{0.00**} & \multicolumn{1}{c}{-11.57} & \multicolumn{1}{c}{0.00**} \\
\hline
\multicolumn{1}{l}{\textbf{Post-crash period}} & \multicolumn{1}{c}{} & \multicolumn{1}{c}{} & \multicolumn{1}{c}{} \\
\hline
$WD_1$ & Commodity & \multicolumn{1}{c}{-7.00}  & \multicolumn{1}{c}{0.00**} & \multicolumn{1}{c}{-13.64} & \multicolumn{1}{c}{0.00**} \\
$WD_1$ & Stock     & \multicolumn{1}{c}{6.31}   & \multicolumn{1}{c}{0.00**} & \multicolumn{1}{c}{-12.66} & \multicolumn{1}{c}{0.00**} \\
$WD_2$ & Commodity & \multicolumn{1}{c}{-13.50} & \multicolumn{1}{c}{0.00**} & \multicolumn{1}{c}{-13.66} & \multicolumn{1}{c}{0.00**} \\
$WD_2$ & Stock     & \multicolumn{1}{c}{12.62}  & \multicolumn{1}{c}{0.00**} & \multicolumn{1}{c}{-13.38} & \multicolumn{1}{c}{0.00**} \\
\hline
\hline
\end{tabular}%
}
\caption{
\justifying \small \textit{Stationarity Test Results for First Differences of $WD_1$ and $WD_2$ time-series across different periods for stock and commodity markets. The table shows statistics and p-values for ADF and PP stationary tests. At a $5\%$ significance level, the time-series are consistently stationary for both markets and across all periods.}}
\label{tab:stationarity_analysis_fd}
\begin{flushleft}
\textit{** indicates significance at the 5\% significance level.}
\end{flushleft}
\end{table}

Table \ref{tab:stationarity_analysis_fd} shows that the p-values of both tests for the first-difference $WD_1$ and $WD_2$ time-series are consistently below the significance level of 0.05 for both markets in all three time periods. Hence, we reject the null hypothesis and, therefore, the first-difference $WD_1$ and $WD_2$ time-series are stationary for stock and commodity markets during all periods. Only the first-difference $WD_2$ time-series for the commodity during the crash is not stationary as per the ADF test. We tested the second-difference $WD_2$ time-series during the crash period and found it to be consistently stationary for both markets at a $5\%$ significance level using both tests.\\

\subsubsection{Stationarity Tests - Sectors}
\label{stationarity_sectors}
Similar to Sec. \ref{stationarity_Stock_commodity}, we divide the time period into three periods: pre-crash, crash, and post-crash. For each sector, we perform stationarity tests using the ADF and PP tests separately for all periods.
The test statistics and corresponding p-values for the stationarity tests for $WD_2$  time-series are given in Table~\ref{tab:stationarity_tests_$WD_2$}.
\begin{table}[h!]
\centering
\resizebox{0.5\textwidth}{!}{%
\begin{tabular}{lrrrr}
\hline
\hline
\hspace{0.3 cm}
\textbf{Sector} & \textbf{ADF Statistic} & \textbf{ADF p-value} & \textbf{PP Statistic} & \textbf{PP p-value} \\
\hline
\multicolumn{1}{l}{\textbf{Pre-crash period}} & \multicolumn{1}{c}{} & \multicolumn{1}{c}{} & \multicolumn{1}{c}{} \\
\hline
Consumer Discretionary  & \multicolumn{1}{c}{-1.62} & \multicolumn{1}{c}{0.47} & \multicolumn{1}{c}{-1.54} & \multicolumn{1}{c}{0.51} \\
Consumer Staples        & \multicolumn{1}{c}{-2.24} & \multicolumn{1}{c}{0.19} & \multicolumn{1}{c}{-2.40} & \multicolumn{1}{c}{0.14} \\
Energy                  & \multicolumn{1}{c}{-2.10} & \multicolumn{1}{c}{0.24} & \multicolumn{1}{c}{-2.21} & \multicolumn{1}{c}{0.20} \\
Financials              & \multicolumn{1}{c}{-1.59} & \multicolumn{1}{c}{0.49} & \multicolumn{1}{c}{-1.90} & \multicolumn{1}{c}{0.33} \\
Healthcare              & \multicolumn{1}{c}{-1.70} & \multicolumn{1}{c}{0.43} & \multicolumn{1}{c}{-1.96} & \multicolumn{1}{c}{0.30} \\
Industrials             & \multicolumn{1}{c}{-1.79} & \multicolumn{1}{c}{0.38} & \multicolumn{1}{c}{-1.66} & \multicolumn{1}{c}{0.45} \\
IT                      & \multicolumn{1}{c}{-1.31} & \multicolumn{1}{c}{0.62} & \multicolumn{1}{c}{-1.45} & \multicolumn{1}{c}{0.56} \\
Materials               & \multicolumn{1}{c}{-1.69} & \multicolumn{1}{c}{0.43} & \multicolumn{1}{c}{-1.63} & \multicolumn{1}{c}{0.47} \\
Real Estate             & \multicolumn{1}{c}{-2.00} & \multicolumn{1}{c}{0.28} & \multicolumn{1}{c}{-2.22} & \multicolumn{1}{c}{0.20} \\
Utilities               & \multicolumn{1}{c}{-4.16} & \multicolumn{1}{c}{0.00**} & \multicolumn{1}{c}{-4.19} & \multicolumn{1}{c}{0.00**} \\
Communications Services & \multicolumn{1}{c}{-2.24} & \multicolumn{1}{c}{0.19} & \multicolumn{1}{c}{-2.61} & \multicolumn{1}{c}{0.09} \\
\hline
\multicolumn{1}{l}{\textbf{Crash period}} & \multicolumn{1}{c}{} & \multicolumn{1}{c}{} & \multicolumn{1}{c}{} \\
\hline
Consumer Discretionary  & \multicolumn{1}{c}{-1.48} & \multicolumn{1}{c}{0.54} & \multicolumn{1}{c}{-1.52} & \multicolumn{1}{c}{0.52} \\
Consumer Staples        & \multicolumn{1}{c}{-2.08} & \multicolumn{1}{c}{0.25} & \multicolumn{1}{c}{-1.52} & \multicolumn{1}{c}{0.52} \\
Energy                  & \multicolumn{1}{c}{-1.23} & \multicolumn{1}{c}{0.66} & \multicolumn{1}{c}{-1.63} & \multicolumn{1}{c}{0.47} \\
Financials              & \multicolumn{1}{c}{-1.95} & \multicolumn{1}{c}{0.31} & \multicolumn{1}{c}{-1.45} & \multicolumn{1}{c}{0.56} \\
Healthcare              & \multicolumn{1}{c}{-1.71} & \multicolumn{1}{c}{0.43} & \multicolumn{1}{c}{-1.75} & \multicolumn{1}{c}{0.41} \\
Industrials             & \multicolumn{1}{c}{-1.88} & \multicolumn{1}{c}{0.34} & \multicolumn{1}{c}{-1.60} & \multicolumn{1}{c}{0.48} \\
IT                      & \multicolumn{1}{c}{-1.62} & \multicolumn{1}{c}{0.48} & \multicolumn{1}{c}{-1.59} & \multicolumn{1}{c}{0.49} \\
Materials               & \multicolumn{1}{c}{-2.28} & \multicolumn{1}{c}{0.18} & \multicolumn{1}{c}{-1.63} & \multicolumn{1}{c}{0.47} \\
Real Estate             & \multicolumn{1}{c}{-1.77} & \multicolumn{1}{c}{0.40} & \multicolumn{1}{c}{-1.43} & \multicolumn{1}{c}{0.57} \\
Utilities               & \multicolumn{1}{c}{-2.33} & \multicolumn{1}{c}{0.16} & \multicolumn{1}{c}{-1.44} & \multicolumn{1}{c}{0.56} \\
Communications Services & \multicolumn{1}{c}{-0.95} & \multicolumn{1}{c}{0.77} & \multicolumn{1}{c}{-1.26} & \multicolumn{1}{c}{0.65} \\
\hline
\multicolumn{1}{l}{\textbf{Post-crash period}} & \multicolumn{1}{c}{} & \multicolumn{1}{c}{} & \multicolumn{1}{c}{} \\
\hline
Consumer Discretionary  & \multicolumn{1}{c}{-2.61} & \multicolumn{1}{c}{0.09} & \multicolumn{1}{c}{-2.06} & \multicolumn{1}{c}{0.26} \\
Consumer Staples        & \multicolumn{1}{c}{-2.31} & \multicolumn{1}{c}{0.17} & \multicolumn{1}{c}{-2.37} & \multicolumn{1}{c}{0.15} \\
Energy                  & \multicolumn{1}{c}{-1.64} & \multicolumn{1}{c}{0.46} & \multicolumn{1}{c}{-1.78} & \multicolumn{1}{c}{0.39} \\
Financials              & \multicolumn{1}{c}{-1.84} & \multicolumn{1}{c}{0.36} & \multicolumn{1}{c}{-2.32} & \multicolumn{1}{c}{0.17} \\
Healthcare              & \multicolumn{1}{c}{-2.11} & \multicolumn{1}{c}{0.24} & \multicolumn{1}{c}{-2.57} & \multicolumn{1}{c}{0.10} \\
Industrials             & \multicolumn{1}{c}{-2.50} & \multicolumn{1}{c}{0.12} & \multicolumn{1}{c}{-2.34} & \multicolumn{1}{c}{0.16} \\
IT                      & \multicolumn{1}{c}{-2.26} & \multicolumn{1}{c}{0.18} & \multicolumn{1}{c}{-2.24} & \multicolumn{1}{c}{0.19} \\
Materials               & \multicolumn{1}{c}{-2.78} & \multicolumn{1}{c}{0.06} & \multicolumn{1}{c}{-2.92} & \multicolumn{1}{c}{0.04**} \\
Real Estate             & \multicolumn{1}{c}{-1.96} & \multicolumn{1}{c}{0.31} & \multicolumn{1}{c}{-2.27} & \multicolumn{1}{c}{0.18} \\
Utilities               & \multicolumn{1}{c}{-0.99} & \multicolumn{1}{c}{0.76} & \multicolumn{1}{c}{-1.33} & \multicolumn{1}{c}{0.62} \\
Communications Services & \multicolumn{1}{c}{-3.08} & \multicolumn{1}{c}{0.03**} & \multicolumn{1}{c}{-2.72} & \multicolumn{1}{c}{0.07} \\
\hline
\hline
\end{tabular}%
}
\caption{\justifying \small \textit{Stationarity Tests for $WD_2$ time-series across different periods for Stock and commodity market. The table shows statistics and p-values for ADF and PP stationary tests. At a $5\%$ significance level, the time-series are not consistently stationary for both tests and across all periods.}}
\label{tab:stationarity_tests_$WD_2$}
\begin{flushleft}
\textit{** indicates significance at a $5\%$ significance level. }
\end{flushleft}
\end{table}

Table \ref{tab:stationarity_tests_$WD_2$} shows the statistics and corresponding p-values for stationarity tests based on the ADF and PP tests. The p-values are significantly greater than the 0.05 significance level. Therefore, we cannot reject the null hypothesis that $WD_2$ time-series are non-stationary. Consequently, we perform the stationarity tests on the first difference $WD_2$ time-series. The results of the tests are listed in Table~\ref{tab:stationarity_tests_$WD_2$_fd}.\\

\begin{table}[h!]
\centering
\resizebox{0.5\textwidth}{!}{%
\begin{tabular}{lrrrr}
\hline
\hline
\textbf{Sector} & \textbf{ADF Statistic} & \textbf{ADF p-value} & \textbf{PP Statistic} & \textbf{PP p-value} \\
\hline
\multicolumn{1}{l}{\textbf{Pre-crash period}} & \multicolumn{1}{c}{} & \multicolumn{1}{c}{} & \multicolumn{1}{c}{} \\
\hline
Consumer Discretionary  & \multicolumn{1}{c}{-12.47} & \multicolumn{1}{c}{0.00**} & \multicolumn{1}{c}{-12.34} & \multicolumn{1}{c}{0.00**} \\
Consumer Staples        & \multicolumn{1}{c}{-14.17} & \multicolumn{1}{c}{0.00**} & \multicolumn{1}{c}{-14.16} & \multicolumn{1}{c}{0.00**} \\
Energy                  & \multicolumn{1}{c}{-13.49} & \multicolumn{1}{c}{0.00**} & \multicolumn{1}{c}{-13.51} & \multicolumn{1}{c}{0.00**} \\
Financials              & \multicolumn{1}{c}{-13.21} & \multicolumn{1}{c}{0.00**} & \multicolumn{1}{c}{-13.15} & \multicolumn{1}{c}{0.00**} \\
Healthcare              & \multicolumn{1}{c}{-14.09} & \multicolumn{1}{c}{0.00**} & \multicolumn{1}{c}{-14.02} & \multicolumn{1}{c}{0.00**} \\
Industrials             & \multicolumn{1}{c}{-13.75} & \multicolumn{1}{c}{0.00**} & \multicolumn{1}{c}{-13.69} & \multicolumn{1}{c}{0.00**} \\
IT                      & \multicolumn{1}{c}{-14.73} & \multicolumn{1}{c}{0.00**} & \multicolumn{1}{c}{-14.74} & \multicolumn{1}{c}{0.00**} \\
Materials               & \multicolumn{1}{c}{-13.90} & \multicolumn{1}{c}{0.00**} & \multicolumn{1}{c}{-13.89} & \multicolumn{1}{c}{0.00**} \\
Real Estate             & \multicolumn{1}{c}{-13.67} & \multicolumn{1}{c}{0.00**} & \multicolumn{1}{c}{-13.63} & \multicolumn{1}{c}{0.00**} \\
Utilities               & \multicolumn{1}{c}{-14.32} & \multicolumn{1}{c}{0.00**} & \multicolumn{1}{c}{-14.31} & \multicolumn{1}{c}{0.00**} \\
Communications Services & \multicolumn{1}{c}{-13.58} & \multicolumn{1}{c}{0.00**} & \multicolumn{1}{c}{-13.52} & \multicolumn{1}{c}{0.00**} \\
\hline
\multicolumn{1}{l}{\textbf{Crash period}} & \multicolumn{1}{c}{} & \multicolumn{1}{c}{} & \multicolumn{1}{c}{} \\
\hline
Consumer Discretionary  & \multicolumn{1}{c}{-7.42}  & \multicolumn{1}{c}{0.00**} & \multicolumn{1}{c}{-14.28} & \multicolumn{1}{c}{0.00**} \\
Consumer Staples        & \multicolumn{1}{c}{-5.56}  & \multicolumn{1}{c}{0.00**} & \multicolumn{1}{c}{-13.52} & \multicolumn{1}{c}{0.00**} \\
Energy                  & \multicolumn{1}{c}{-10.39} & \multicolumn{1}{c}{0.00**} & \multicolumn{1}{c}{-14.25} & \multicolumn{1}{c}{0.00**} \\
Financials              & \multicolumn{1}{c}{-6.28}  & \multicolumn{1}{c}{0.00**} & \multicolumn{1}{c}{-13.45} & \multicolumn{1}{c}{0.00**} \\
Healthcare              & \multicolumn{1}{c}{-8.11}  & \multicolumn{1}{c}{0.00**} & \multicolumn{1}{c}{-14.13} & \multicolumn{1}{c}{0.00**} \\
Industrials             & \multicolumn{1}{c}{-7.34}  & \multicolumn{1}{c}{0.00**} & \multicolumn{1}{c}{-13.91} & \multicolumn{1}{c}{0.00**} \\
IT                      & \multicolumn{1}{c}{-5.38}  & \multicolumn{1}{c}{0.00**} & \multicolumn{1}{c}{-12.23} & \multicolumn{1}{c}{0.00**} \\
Materials               & \multicolumn{1}{c}{-6.95}  & \multicolumn{1}{c}{0.00**} & \multicolumn{1}{c}{-13.67} & \multicolumn{1}{c}{0.00**} \\
Real Estate             & \multicolumn{1}{c}{-7.22}  & \multicolumn{1}{c}{0.00**} & \multicolumn{1}{c}{-13.92} & \multicolumn{1}{c}{0.00**} \\
Utilities               & \multicolumn{1}{c}{-6.89}  & \multicolumn{1}{c}{0.00**} & \multicolumn{1}{c}{-13.80} & \multicolumn{1}{c}{0.00**} \\
Communications Services & \multicolumn{1}{c}{-5.61}  & \multicolumn{1}{c}{0.00**} & \multicolumn{1}{c}{-13.37} & \multicolumn{1}{c}{0.00**} \\
\hline
\multicolumn{1}{l}{\textbf{Post-crash period}} & \multicolumn{1}{c}{} & \multicolumn{1}{c}{} & \multicolumn{1}{c}{} \\
\hline
Consumer Discretionary  & \multicolumn{1}{c}{-9.12}  & \multicolumn{1}{c}{0.00**} & \multicolumn{1}{c}{-14.18} & \multicolumn{1}{c}{0.00**} \\
Consumer Staples        & \multicolumn{1}{c}{-9.06}  & \multicolumn{1}{c}{0.00**} & \multicolumn{1}{c}{-14.34} & \multicolumn{1}{c}{0.00**} \\
Energy                  & \multicolumn{1}{c}{-7.78}  & \multicolumn{1}{c}{0.00**} & \multicolumn{1}{c}{-14.67} & \multicolumn{1}{c}{0.00**} \\
Financials              & \multicolumn{1}{c}{-8.97}  & \multicolumn{1}{c}{0.00**} & \multicolumn{1}{c}{-13.92} & \multicolumn{1}{c}{0.00**} \\
Healthcare              & \multicolumn{1}{c}{-10.23} & \multicolumn{1}{c}{0.00**} & \multicolumn{1}{c}{-14.07} & \multicolumn{1}{c}{0.00**} \\
Industrials             & \multicolumn{1}{c}{-9.82}  & \multicolumn{1}{c}{0.00**} & \multicolumn{1}{c}{-13.94} &  \multicolumn{1}{c}{0.00**} \\
Materials               & \multicolumn{1}{c}{-9.90}  & \multicolumn{1}{c}{0.00**} & \multicolumn{1}{c}{-14.23} & \multicolumn{1}{c}{0.00**} \\
Real Estate             & \multicolumn{1}{c}{-10.12} & \multicolumn{1}{c}{0.00**} & \multicolumn{1}{c}{-14.16} & \multicolumn{1}{c}{0.00**} \\
Utilities               & \multicolumn{1}{c}{-11.00} & \multicolumn{1}{c}{0.00**} & \multicolumn{1}{c}{-14.19} & \multicolumn{1}{c}{0.00**} \\
Communications Services & \multicolumn{1}{c}{-10.53} & \multicolumn{1}{c}{0.00**} & \multicolumn{1}{c}{-14.07} & \multicolumn{1}{c}{0.00**} \\
\hline
\hline
\end{tabular}%
}
\caption{\justifying \small
\textit{Stationarity tests for first differences of $WD_2$ time-series across different periods for various sectors. The table shows statistics and p-values for ADF and PP stationary tests. At a $5\%$ significance level, the time-series are consistently stationary for both tests and across all periods.}}
\label{tab:stationarity_tests_$WD_2$_fd}
\begin{flushleft}
\textit{** indicates significance at the $5\%$ significance level.}
\end{flushleft}
\end{table}

Table \ref{tab:stationarity_tests_$WD_2$_fd} shows the statistics and corresponding p-values for stationarity tests based on ADF and PP tests. The p-values are significantly less than the $0.05$ significance level. Therefore, the null hypothesis is rejected and the first difference $WD_2$ series is stationary for all sectors for all time periods.

\subsection{Causal Effect}
In Sec.~\ref{stationarity_Stock_commodity} and \ref{stationarity_sectors}, we showed that the $WD$ time-series for stock, commodity, and sectoral markets are non-stationary for the three periods. However, the corresponding first difference time-series is stationary for all the time periods. In this section, we perform the Granger-causality tests between the first-difference $WD$ time-series of stock and commodity markets for pre-crash, crash, and post-crash periods separately. Moreover, a pairwise Granger-causality test between different sectors is performed for the three periods. The inference of the Granger-causality test is derived from the value of the F-statistic and its corresponding p-value. For two time-series X and Y, the null hypothesis states that "X does not Granger-cause Y". In this study, we are considering the 0.05 significance level. Therefore, if the p-value is less than 0.05, then we infer that the time-series X does Granger-cause time-series Y. A separate test is performed to check whether time-series Y Granger causes time-series X. If both are are Granger-causing each other, then the relation is said to be \emph{bidirectional} Granger-causal relation. If, however, only one of the time-series Granger-causes the other but the latter time-series does not Granger-cause the former time-series, then the relation is said to be \emph{unidirectional} Granger-causal relation. If both time-series do not Granger-cause each other, then the two time-series are termed \emph{independent}~\cite{guo2010granger}.

\subsubsection{Stock and Commodity}
\label{causality_Stock_comm}

In this Section, we study the Granger-causal relation, if any, between the topological dynamics of stock and commodity markets during pre-crash, crash, and post-crash periods using the first-difference $WD_1$ and $WD_2$ time-series. We test the Granger-causal relation between the corresponding first-difference $WD_1$ and $WD_2$ time-series of stock and commodity market separately for the three periods. The test results are listed in Table~\ref{tab:Granger_causality_analysis}.\\
\begin{table}[h!]
\centering
\resizebox{0.5\textwidth}{!}{%
\begin{tabular}{llcrr}
\hline
\hline
\textbf{Metric} & \textbf{Direction} & \textbf{Optimal Lag (FPE)} & \textbf{F-Statistic} & \textbf{P-Value} \\
\hline
\multicolumn{1}{l}{\textbf{Pre-crash period}} & \multicolumn{1}{c}{} & \multicolumn{1}{c}{} & \multicolumn{1}{c}{} \\
\hline
$WD_1$ & \multicolumn{1}{c}{Stock → Commodity} & \multicolumn{1}{c}{5} & \multicolumn{1}{c}{3.22} & \multicolumn{1}{c}{0.01**} \\
$WD_1$ & \multicolumn{1}{c}{Commodity → Stock} & \multicolumn{1}{c}{5} & \multicolumn{1}{c}{0.31} & \multicolumn{1}{c}{0.91} \\
$WD_2$ & \multicolumn{1}{c}{Stock → Commodity} & \multicolumn{1}{c}{1} & \multicolumn{1}{c}{2.48} & \multicolumn{1}{c}{0.12} \\
$WD_2$ & \multicolumn{1}{c}{Commodity → Stock} & \multicolumn{1}{c}{1} & \multicolumn{1}{c}{0.14} & \multicolumn{1}{c}{0.71} \\
\hline
\multicolumn{1}{l}{\textbf{Crash period}} & \multicolumn{1}{c}{} & \multicolumn{1}{c}{} & \multicolumn{1}{c}{} \\
\hline
$WD_1$ & \multicolumn{1}{c}{Stock → Commodity} & \multicolumn{1}{c}{5} & \multicolumn{1}{c}{2.92} & \multicolumn{1}{c}{0.01**} \\
$WD_1$ & \multicolumn{1}{c}{Commodity → Stock} & \multicolumn{1}{c}{5} & \multicolumn{1}{c}{8.37} & \multicolumn{1}{c}{0.00**} \\
$WD_2$ & \multicolumn{1}{c}{Stock → Commodity} & \multicolumn{1}{c}{5} & \multicolumn{1}{c}{2.87} & \multicolumn{1}{c}{0.02**} \\
$WD_2$ & \multicolumn{1}{c}{Commodity → Stock} & \multicolumn{1}{c}{5} & \multicolumn{1}{c}{8.56} & \multicolumn{1}{c}{0.00**} \\
\hline
\multicolumn{1}{l}{\textbf{Post-crash period}} & \multicolumn{1}{c}{} & \multicolumn{1}{c}{} & \multicolumn{1}{c}{} \\
\hline
$WD_1$ & \multicolumn{1}{c}{Stock → Commodity} & \multicolumn{1}{c}{3} & \multicolumn{1}{c}{5.04} & \multicolumn{1}{c}{0.00**} \\
$WD_1$ & \multicolumn{1}{c}{Commodity → Stock} & \multicolumn{1}{c}{3} & \multicolumn{1}{c}{1.03} & \multicolumn{1}{c}{0.38} \\
$WD_2$ & \multicolumn{1}{c}{Stock → Commodity} & \multicolumn{1}{c}{3} & \multicolumn{1}{c}{4.18} & \multicolumn{1}{c}{0.01**} \\
$WD_2$ & \multicolumn{1}{c}{Commodity → Stock} & \multicolumn{1}{c}{3} & \multicolumn{1}{c}{1.18} & \multicolumn{1}{c}{0.32} \\
\hline
\hline
\end{tabular}%
}
\caption{\justifying \small
\textit{Granger-causality analysis across different periods. The results at 5\% significance level show that both before and after the market crash due to COVID-19, the stock market was dominant unidirectionally. However, in the year leading up to the crash and during the crash, both the markets influenced each other.}}
\label{tab:Granger_causality_analysis}
\begin{flushleft}
\textit{** indicates significance at the 5\% significance level.}
\end{flushleft}
\end{table}

Table~\ref{tab:Granger_causality_analysis} shows the F-statistic and p-value for the Granger-causality tests between the $WD$ time-series of stock and commodity markets during the three periods. For the pre-crash period, the p-values for causality tests from stock to commodity market are less than the 0.05 significance level for $WD_1$ and 0.12 for $WD_2$. However, the corresponding p-values for  Granger-causality from commodity to stock market are significantly higher than the 0.05 significance level. Therefore, the change in topological dynamics of the stock market influences the same of the commodity market showing the general influence of the stock market over the commodity market during a normal period. For the crash period, the p-values for Granger-causality tests for both directions are consistently below the $5\%$ significance level. Hence, the change in topological dynamics of the two markets influences each other during the crash period. The test was repeated for the second difference $WD_2$ time-series as the first-difference time-series was not consistently stationary as highlighted in Table~\ref{tab:stationarity_analysis_fd}. A similar result was obtained for the second difference $WD_2$ time-series. Our result corroborates with works that have shown a greater bidirectional return and volatility
spillovers across stock-commodity markets during the COVID-19 outbreak~\cite{liu2022dynamic}. For the post-crash period, the p-values for the Granger-causality from stock to commodity market are consistently below the 0.05 significance level. However, the corresponding p-values for Granger-causality from commodity to stock market are significantly above the 0.05 significance level. Therefore, the change in topological dynamics of the stock market influences the same of the commodity market. This highlights the general dominance of the stock market over the commodity market.

%the dominnace of the Stock   markets form a Granger-causal relation in all three periods analyzed. We observe that during the pre- and post-crash periods, the topological change in the  Stock market is Granger-causing the same in the commodity market, which demonstrates the general dominance of the Stock market over the commodity market (\textcolor{red}{explain with values of F and P}). However, during the crash period, the topological dynamics of the Stock and commodity markets form a bi-directional Granger-causal relation. This signifies the general interconnection and interdependence in the two markets during the crisis period. Moreover, 

%This may be due to the nature of crash due to COVID-19, wherein the supply chains of the commodities were disrupted directly by COVID-19 and measures
%taken to contain it~\cite{bank2020shock}. 

%This suggests that during the normal time period, Stock market influences the commodity market. However, during the pandemic like the Covid-19 which effected the movement of commodities before the stock market [cite], the commodity market is dominant/influences over the Stock market. However, in the recovery period, i.e., the post-crash period, the Stock market drives the recovery showing the general dominance/influence of Stock market over commodity market.

\subsubsection{Sectors}

\begin{figure}[H]
    \centering
    \includegraphics[width=0.9\linewidth]{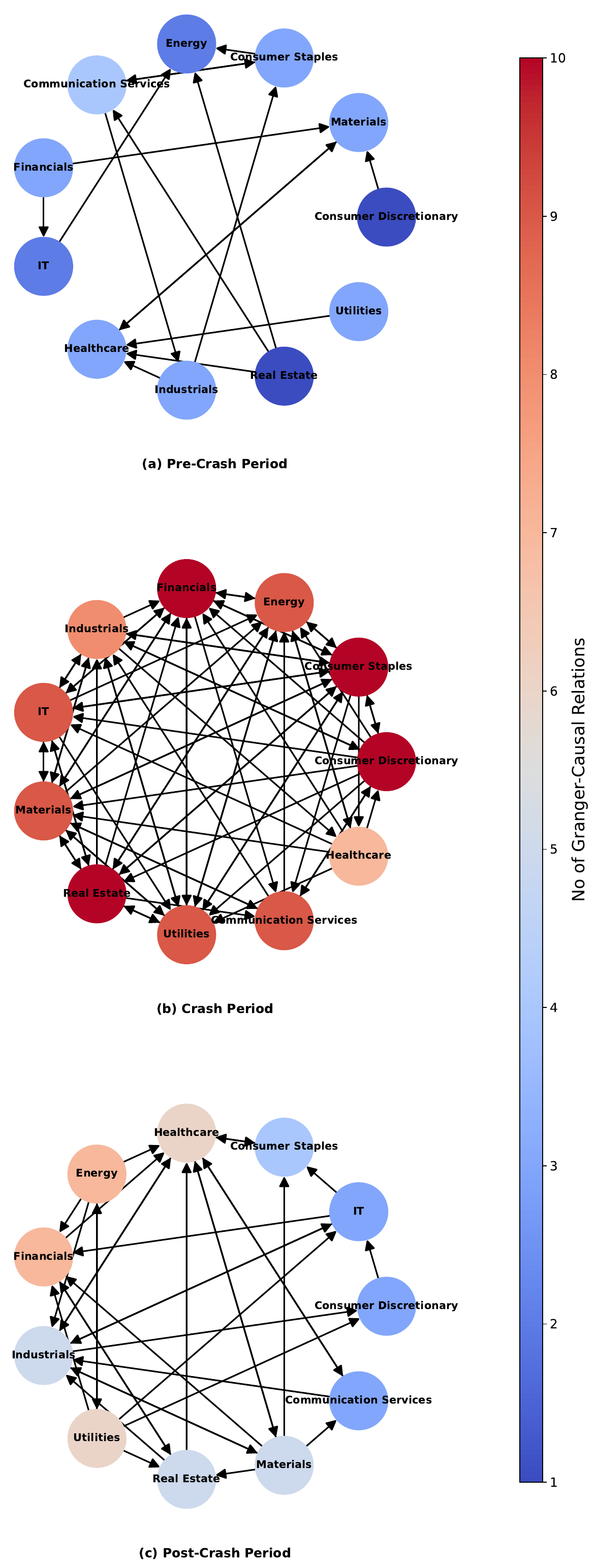}
    \caption{\justifying \small 
    \textit{
    Network graph of the US sectors for different time periods.  Fig. (a) represents the network graph during the pre-crash period, Fig. (b) represents the network graph during the crash period and Fig. (c) represents the network graph in the post-crash period.}}
    \label{fig:network}
\end{figure}

As the first-difference $WD_2$ time-series is stationary for all sectors and all time-periods, we perform the pair-wise Granger-causality tests between the first-difference $WD_2$ time-series of each sector. Tests are conducted separately for pre-crash, crash, and post-crash periods. The results obtained are shown in Table~\ref{tab:causality_analysis} and Fig.~\ref{fig:network}.\\
\begin{table}[h!]
\centering
\resizebox{0.5\textwidth}{!}{%
\begin{tabular}{lrrr}
\hline
\hline
\textbf{Sector} & \textbf{Unidirectional Cause} & \textbf{Unidirectional Effect} & \textbf{Bidirectional Causality} \\
\hline
\multicolumn{1}{l}{\textbf{Pre-crash period}} & \multicolumn{1}{c}{} & \multicolumn{1}{c}{} & \multicolumn{1}{c}{} \\
\hline
Consumer Discretionary & \multicolumn{1}{c}{1} & \multicolumn{1}{c}{0} & \multicolumn{1}{c}{0} \\
Consumer Staples       & \multicolumn{1}{c}{1} & \multicolumn{1}{c}{1} & \multicolumn{1}{c}{1} \\
Energy                 & \multicolumn{1}{c}{0} & \multicolumn{1}{c}{3} & \multicolumn{1}{c}{0} \\
Financials             & \multicolumn{1}{c}{2} & \multicolumn{1}{c}{0} & \multicolumn{1}{c}{0} \\
Healthcare             & \multicolumn{1}{c}{0} & \multicolumn{1}{c}{3} & \multicolumn{1}{c}{1} \\
Industrials            & \multicolumn{1}{c}{2} & \multicolumn{1}{c}{1} & \multicolumn{1}{c}{0} \\
IT                     & \multicolumn{1}{c}{1} & \multicolumn{1}{c}{1} & \multicolumn{1}{c}{0} \\
Materials              & \multicolumn{1}{c}{0} & \multicolumn{1}{c}{2} & \multicolumn{1}{c}{1} \\
Real Estate            & \multicolumn{1}{c}{3} & \multicolumn{1}{c}{0} & \multicolumn{1}{c}{0} \\
Utilities              & \multicolumn{1}{c}{1} & \multicolumn{1}{c}{0} & \multicolumn{1}{c}{0} \\
Communications Services & \multicolumn{1}{c}{1} & \multicolumn{1}{c}{1} & \multicolumn{1}{c}{1} \\
\hline
\multicolumn{1}{l}{\textbf{Crash period}} & \multicolumn{1}{c}{} & \multicolumn{1}{c}{} & \multicolumn{1}{c}{} \\
\hline
Consumer Discretionary & \multicolumn{1}{c}{6} & \multicolumn{1}{c}{1} & \multicolumn{1}{c}{3} \\
Consumer Staples       & \multicolumn{1}{c}{2} & \multicolumn{1}{c}{0} & \multicolumn{1}{c}{8} \\
Energy                 & \multicolumn{1}{c}{0} & \multicolumn{1}{c}{4} & \multicolumn{1}{c}{5} \\
Financials             & \multicolumn{1}{c}{1} & \multicolumn{1}{c}{4} & \multicolumn{1}{c}{5} \\
Healthcare             & \multicolumn{1}{c}{5} & \multicolumn{1}{c}{2} & \multicolumn{1}{c}{1} \\
Industrials            & \multicolumn{1}{c}{2} & \multicolumn{1}{c}{2} & \multicolumn{1}{c}{5} \\
IT                     & \multicolumn{1}{c}{2} & \multicolumn{1}{c}{2} & \multicolumn{1}{c}{5} \\
Materials              & \multicolumn{1}{c}{1} & \multicolumn{1}{c}{2} & \multicolumn{1}{c}{7} \\
Real Estate            & \multicolumn{1}{c}{3} & \multicolumn{1}{c}{1} & \multicolumn{1}{c}{5} \\
Utilities              & \multicolumn{1}{c}{0} & \multicolumn{1}{c}{3} & \multicolumn{1}{c}{6} \\
Communications Services & \multicolumn{1}{c}{2} & \multicolumn{1}{c}{3} & \multicolumn{1}{c}{2} \\
\hline
\multicolumn{1}{l}{\textbf{Post-crash period}} & \multicolumn{1}{c}{} & \multicolumn{1}{c}{} & \multicolumn{1}{c}{} \\
\hline
Consumer Discretionary & \multicolumn{1}{c}{1} & \multicolumn{1}{c}{2} & \multicolumn{1}{c}{0} \\
Consumer Staples       & \multicolumn{1}{c}{0} & \multicolumn{1}{c}{2} & \multicolumn{1}{c}{1} \\
Energy                 & \multicolumn{1}{c}{3} & \multicolumn{1}{c}{0} & \multicolumn{1}{c}{1} \\
Financials             & \multicolumn{1}{c}{1} & \multicolumn{1}{c}{4} & \multicolumn{1}{c}{1} \\
Healthcare             & \multicolumn{1}{c}{0} & \multicolumn{1}{c}{3} & \multicolumn{1}{c}{4} \\
Industrials            & \multicolumn{1}{c}{1} & \multicolumn{1}{c}{3} & \multicolumn{1}{c}{3} \\
IT                     & \multicolumn{1}{c}{2} & \multicolumn{1}{c}{2} & \multicolumn{1}{c}{1} \\
Materials              & \multicolumn{1}{c}{4} & \multicolumn{1}{c}{0} & \multicolumn{1}{c}{2} \\
Real Estate            & \multicolumn{1}{c}{2} & \multicolumn{1}{c}{2} & \multicolumn{1}{c}{1} \\
Utilities              & \multicolumn{1}{c}{4} & \multicolumn{1}{c}{0} & \multicolumn{1}{c}{1} \\
Communications Services & \multicolumn{1}{c}{1} & \multicolumn{1}{c}{1} & \multicolumn{1}{c}{1} \\
\hline
\hline
\end{tabular}%
}
\caption{\justifying \small
\textit{Pairwise Granger-causality analysis of sectors across different periods. The table shows the number of unidirectional causes, unidirectional effects, and bidirectional causalities for each sector across the pre-crash, crash, and post-crash periods. A significant increase in the Granger-causal relations is observed between the US sectors during the crash period.}}
\label{tab:causality_analysis}
\end{table}

Table \ref{tab:causality_analysis} represents the counts of uni-directional cause, uni-directional effect, and bi-directional Granger-causal relations for each sector during the different time periods. We observe considerably less number of Granger-causal relations (3 per sector) between the sectors in the pre-crash period. Also, the total number of bidirectional causal relations between the sectors is only 2. During the crash period, there is a significant increase in the Granger-causal relations between the sectors (9 per sector), signifying greater inter-dependence among the different sectors during the crash period. Also, the number of bidirectional Granger-causal relations has risen considerably during this period to 26. In the post-crash period, there is a general decline in the Granger-causal relations between the sectors (5 per sector). However, the number of Granger-causal relations is greater than in the pre-crash period. Moreover, the bidirectional causal relations after the crash decrease to $8$, which is greater than the pre-crash period. This signifies the relatively volatile nature of the market due to recurrent COVID-19 waves and aftershocks. 
Figs.~\ref{fig:network}a-~\ref{fig:network}c shows the network graph of the US Stock sectors for three time periods.  Fig.~\ref{fig:network}(a) represents the network graph during the pre-crash period, Fig.~\ref{fig:network}(b) represents the network graph during the crash period and Fig.~\ref{fig:network}(c) represents the network graph in the post-crash period. Each node represents the sectors, and their color represents the Granger-causal behavior of the sectors, respectively. The behavior is derived by adding each unique Granger-causal relation for each sector. Sectors on the reddish side have more Granger-causal relations and those on the bluish side have fewer Granger-causal relations. The connections between the sectors represent the Granger-causal relationship between them and the direction of the arrow represents the direction of Granger-causality. We observe a significant increase in the Granger-causal relations between the sectors during the crash period. The relations decrease after the crash but remain greater than in the pre-crash period.

\section{Conclusion}
\label{con}
In this work, we have detected the crashes in the stock and commodity markets due to the COVID-19 pandemic using tools from topological data analysis (TDA). The use of TDA for the identification of the crash for the whole market is deemed ideal as it can handle multidimensional time-series simultaneously and reflect the collective dynamics of the market as a whole. The Wasserstein distance ($WD$), a metric comparing topologies of point clouds, is used to identify crashes. $WD$ spikes abruptly during the crash period, indicating a drastic change in the topology. In addition, we compared the stock and commodity markets and observed significant differences in their topologies during the crash periods. This could be due to two reasons 1) Difference in the magnitude of the crash and/or 2) Temporal lag in the topological changes between the two markets. We obtained that the topological magnitudes of the crash are comparable for the two markets. In addition, the Granger causality test was carried out to identify any causal relationship between the two markets.

Granger-causal analyses were carried out for three periods, viz. pre-crash, crash, and post-crash, each spanning a year. During the normal period (pre-crash and post-crash) we found that the stock market is dominant over the commodity market. However, during the crash, both markets influence each other. During the crash period, both the market moves in synchronization where the movement of one influences the other. The results agree with the existing literature~\cite{liu2022dynamic}.

Similar analyses were carried out for different sectors of the US stock market. The crash due to COVID-19 was identified using TDA for different sectors. Topological comparisons of different sectors were also performed. We observed that the sectoral topology was significantly different during the crash period which was visualized by the significant spike in $WD$. Hence, we compare the topological magnitudes of the crash for different sectors. To check for causation, pair-wise Granger-causality tests were carried out between the sectors for the pre-crash, crash, and post-crash periods. The pre-crash period shows a considerably low number of Granger-causal relations between the sectors. This signifies that during the normal period, the sectors moved independently based on their own fundamentals and future outlook. However, during the crash period, the number of causal relations shows a significant increase for all sectors showing greater interdependence of the sectors during the crash period. This clearly shows that during the crash period, the sectors move in a synchronized manner where the movement of one sector impacts the movement of others. The post-crash period shows a general decrease in the number of Granger causal relations between the sectors. This indicates that the market remains relatively interdependent in the post-crash period as compared to the pre-crash period.

The paper successfully studies the dynamics of the markets and sectors using the robust technique of TDA and the application of Granger-causality properly analyses the interdependencies between the different markets and sectors. Understanding the dynamics of different markets and sectors and analyzing their evolution of relation during different market periods is of utmost importance as different markets and sectors follow relatively different dynamics during normal periods. The increase in interdependence between different sectors or markets could act as an indicator of a possible crash thus helping investors to make informed decisions regarding their investments.

\begin{acknowledgments}
We would like to acknowledge NIT Sikkim for allocating doctoral fellowship to Buddha Nath Sharma and SR Luwang. We also acknowledge the inputs provided by Kundan Mukhia. 
\end{acknowledgments}

\section*{Data Availability Statement}

The data that support the findings of this study are openly available on the Yahoo Finance website at \url{https://finance.yahoo.com}~\cite{Yahoo}.

\section*{REFERENCES}
\bibliographystyle{unsrt}
\nocite{*}
\bibliography{References}   

%\clearpage
\appendix

\section{List of stocks and commodities}

\label{appendix}

\begin{table}[H]
\centering
\begin{tabular}{|l|l|}
\hline
\textbf{Category} & \textbf{Stocks/Commodities Taken} \\ \hline
\textbf{US Stocks} & Apple (AAPL), Microsoft (MSFT),\\& Alphabet (GOOGL), Amazon (AMZN),\\& Nvidia (NVDA), Meta (META),\\&Tesla (TSLA), Visa (V),\\& Berkshire Hathaway (BRK-B),\\&  UnitedHealth (UNH), Eli Lilly (LLY),\\& Johnson \& Johnson (JNJ),\\& ExxonMobil (XOM), Walmart (WMT),\\& Procter \& Gamble (PG), Mastercard (MA),\\& Chevron (CVX), Broadcom (AVGO),\\& Home Depot (HD), Merck (MRK) \\ \hline
\textbf{Commodities} & Crude Oil (CL=F), Brent Crude (BZ=F),\\& Gasoline (RB=F), Heating Oil (HO=F),\\& Natural Gas (NG=F), Aluminum (ALI=F),\\& Copper (HG=F), Zinc (ZN=F),\\& Gold (GC=F), Silver (SI=F),\\& Corn (ZC=F), Wheat (ZW=F),\\& Kansas Wheat (KE=F), Soybeans (ZS=F),\\& Cotton (CT=F), Coffee (KC=F),\\& Cocoa (CC=F), Live Cattle (LE=F),\\& Feeder Cattle (GF=F), Lean Hogs (HE=F) \\ \hline
\end{tabular}
\caption{\justifying \small \textit{The table represents the list of stocks and commodities used in the analysis. The ticker of each stock and commodity is given in the bracket.}}
\label{tab:stocks_commodities}
\end{table}

\begin{table}[H]
\centering
\small
\begin{tabular}{|l|l|}
\hline
\textbf{Sector} & \textbf{Stocks Taken} \\ \hline
Consumer Discretionary & Amazon (AMZN), Tesla (TSLA), \\& Home Depot (HD), McDonald's (MCD), \\& Nike (NKE) \\ \hline
Consumer Staples & Walmart (WMT), Procter \& Gamble (PG), \\& Costco (COST), Coca-Cola (KO), \\& PepsiCo (PEP) \\ \hline
Energy & ExxonMobil (XOM), Chevron (CVX), \\& ConocoPhillips (COP), Schlumberger (SLB), \\& EOG Resources (EOG) \\ \hline
Financials & Berkshire Hathaway (BRK-B), Visa (V), \\& JPMorgan Chase (JPM), Mastercard (MA), \\& Bank of America (BAC) \\ \hline
Healthcare & Eli Lilly (LLY), UnitedHealth (UNH), \\& Johnson \& Johnson (JNJ), Merck (MRK), \\& AbbVie (ABBV) \\ \hline
Industrials & Union Pacific (UNP), Caterpillar (CAT), \\& General Electric (GE),\\& United Parcel Service (UPS), \\& Honeywell (HON) \\ \hline
IT & Microsoft (MSFT), Apple (AAPL), \\& Nvidia (NVDA), Broadcom (AVGO), \\& Oracle (ORCL) \\ \hline
Materials & Linde (LIN), Sherwin-Williams (SHW), \\& Southern Copper (SCCO),\\& Air Products and Chemicals (APD), \\& Ecolab (ECL) \\ \hline
Real Estate & Prologis (PLD), American Tower (AMT), \\& Equinix (EQIX), Simon Property Group (SPG), \\& Public Storage (PSA) \\ \hline
Utilities & NextEra Energy (NEE),\\& Southern Company (SO), \\& Duke Energy (DUK), Sempra Energy (SRE), \\& American Electric Power (AEP) \\ \hline
Communications Services & Alphabet (GOOGL), Meta Platforms (META), \\& Netflix (NFLX), T-Mobile (TMUS), \\& Comcast (CMCSA) \\ \hline
\end{tabular}
\caption{\justifying \small \textit{The table represents the stocks taken from different sectors of the USA. The ticker of each stock is given in the bracket.}}
\label{tab:us_sectors_stocks}
\end{table}
%\nocite{*}
%\bibliography{References}% Produces the bibliography via BibTeX.

\end{document}